%% file: main.tex
\documentclass[10pt,twoside]{article}
\usepackage[letterpaper,top=0.87in,bottom=0.87in,left=0.75in,right=0.75in]{geometry}
\usepackage[T1]{fontenc}
\usepackage[utf8]{inputenc}
\usepackage{newtxtext}
\usepackage[scaled=0.92]{helvet}
\usepackage[protrusion=true,expansion=true]{microtype}
\usepackage{amsmath}
\usepackage{amssymb}
\usepackage{mathtools}
\usepackage{newtxmath}
\usepackage{siunitx}
\DeclareSIUnit{\molar}{M}
\DeclareSIUnit{\kcal}{kcal}
\DeclareSIUnit{\angstrom}{\text{Å}}
\DeclareSIUnit{\calorie}{cal}
\DeclareSIUnit\bar{bar}
\usepackage{float}
\usepackage{graphicx}
\usepackage{booktabs}
\usepackage{tabularx}
\usepackage{multirow}
\usepackage{makecell}
\usepackage{subfig}
\usepackage[table]{xcolor}
\usepackage[most]{tcolorbox}
\usepackage[font=small,labelfont={sf,bf}]{caption}
\usepackage{tikz}
\definecolor{beadC}{HTML}{628B28}
\definecolor{beadN}{HTML}{1EA69C}
\definecolor{beadP}{HTML}{3362B0}
\definecolor{beadQ}{HTML}{CC3164}
\definecolor{beadX}{HTML}{F78746}
\definecolor{beadlabel}{HTML}{8B1A1A}
\tikzset{
  cgpanel/.style={x=0.2555cm, y=0.2555cm, line cap=round,
                  line join=round},
  cghull/.style={fill opacity=0.3, draw opacity=0.9, line width=0.4pt},
  cghulldash/.style={dash pattern=on 1.2pt off 0.9pt},
  cgbond/.style={line width=0.55pt, draw=black!85},
  cgatom/.style={inner sep=0.4pt, fill=white,
                 font=\fontsize{5.0}{5.0}\selectfont},
  cgcharge/.style={inner sep=0pt,
                   font=\fontsize{3.0}{3.0}\selectfont},
  cglabel/.style={inner sep=0.4pt, text=beadlabel,
                  font=\fontsize{4.0}{4.0}\selectfont},
  cgleader/.style={line width=0.25pt, draw=beadlabel},
  cgkey/.style={fill opacity=0.55, draw opacity=0.9, line width=0.4pt},
}

\usepackage{titlesec}
\titleformat{\section}{\sffamily\bfseries\large}{\thesection}{0.6em}{}
\titleformat{\subsection}{\sffamily\bfseries\normalsize}{\thesubsection}{0.6em}{}
\titleformat{\subsubsection}{\sffamily\itshape\normalsize}{\thesubsubsection}{0.6em}{}
\titlespacing*{\section}{0pt}{1.1\baselineskip}{0.35\baselineskip}
\titlespacing*{\subsection}{0pt}{0.8\baselineskip}{0.25\baselineskip}
\titlespacing*{\subsubsection}{0pt}{0.6\baselineskip}{0.2\baselineskip}
\usepackage{titling}
\pretitle{\begingroup\raggedright\sffamily\bfseries\LARGE}
\posttitle{\par\endgroup\vskip 1em}
\preauthor{\begingroup\raggedright}
\postauthor{\par\endgroup}
\usepackage{authblk}

\usepackage{balance}
\usepackage[numbers,sort&compress,super]{natbib}

\usepackage{xurl}
\usepackage[colorlinks=true,allcolors=blue!55!black,breaklinks=true]{hyperref}
\usepackage[capitalise,noabbrev]{cleveref}
\providecommand{\doi}[1]{\href{https://doi.org/#1}{doi:~\nolinkurl{#1}}}
\makeatletter
\@ifundefined{LaTeXML}{\newif\ifltxml\ltxmlfalse}{\newif\ifltxml\ltxmltrue}
\makeatother

\newlength{\aboxmargin}
\newtcolorbox{abstractbox}{
  colback   = black!4,
  colframe  = black!4,
  arc       = 2mm,
  boxrule   = 0pt,
  top = 4mm, bottom = 4mm,
  left      = \aboxmargin, right = \aboxmargin,
  width     = \dimexpr\textwidth+2\aboxmargin+2.0mm\relax,
  before    = \noindent\hspace*{\dimexpr-\aboxmargin-1mm\relax},
}

\title{
  Predictable Accuracy of Martini 3 Absolute Protein--Ligand Binding Free Energies across 15 Targets
}

\ifltxml
  \author[1,a]{Luis J. Walter}
  \author[1,2,b]{Tristan Bereau}
  \affil[1]{Institute for Theoretical Physics, Heidelberg University, Heidelberg, Germany}
  \affil[2]{Interdisciplinary Center for Scientific Computing (IWR), Heidelberg University, Heidelberg, Germany}
  \affil[a]{Email: \href{mailto:walter@thphys.uni-heidelberg.de}{walter@thphys.uni-heidelberg.de}}
  \affil[b]{Email: \href{mailto:bereau@uni-heidelberg.de}{bereau@uni-heidelberg.de}}
\else
  \author[1]{Luis J. Walter}
  \author[1,2]{Tristan Bereau}
  \affil[1]{Institute for Theoretical Physics, Heidelberg University, Heidelberg, Germany}
  \affil[2]{Interdisciplinary Center for Scientific Computing (IWR), Heidelberg University, Heidelberg, Germany}
  \affil[{ }]{Emails: \href{mailto:walter@thphys.uni-heidelberg.de}{walter@thphys.uni-heidelberg.de}, \href{mailto:bereau@uni-heidelberg.de}{bereau@uni-heidelberg.de}}
\fi
\date{}

\begin{document}

\newcommand{\abstracttext}{%
  Alchemical free-energy calculations predict protein--ligand binding affinities with close-to-experimental accuracy, but their cost restricts them to small sets of ligands.
  Coarse-grained (CG) models such as Martini~3 retain a rigorous statistical-mechanical treatment at a fraction of this cost, yet their accuracy for protein--ligand binding has not been assessed systematically.
  Here, we compute absolute binding free energies for 172 ligands across 15 protein targets with a Martini~3 double-decoupling protocol and compare them against atomistic FEP+ and OpenFE references.
  Martini underbinds for every target, by \SI{2.7}{} to \SI{8.7}{\kilo\calorie\per\mole}.
  This offset is systematic: it comprises a uniform contribution that grows with the number of charged pocket residues.
  A linear model in this number, which requires only the protein structure, reduces the mean absolute error from \SI{5.7}{} to \SI{1.2}{\kilo\calorie\per\mole} in leave-one-out cross-validation, close to the \SI{1.0}{\kilo\calorie\per\mole} obtained with exact per-target offsets.
  The ranking of ligands within a target varies more widely (median Spearman $\rho$ of 0.32, from $-0.41$ to 0.89) and is rationalized by the structure--activity relationship of each series: Martini ranks ligands well when their affinity is driven by hydrophobicity, $\pi$-stacking, or size, and poorly when it is driven by hydrogen bonding or charge.
  Both trends follow from Martini's calibration against bulk partitioning, which captures neither the dielectric environment of a binding pocket nor the directionality of hydrogen bonds.
  While MMGBSA retains the atomistic detail needed to rank congeneric ligands, Martini yields physically meaningful free energies, and its coarse-graining maps many atomistically distinct molecules onto one representation, making it a complementary tool for rapidly spanning chemical space.
}

\ifltxml
  \maketitle
  \begin{abstract}
    \abstracttext
  \end{abstract}
\else
  \makeatletter
  \twocolumn[
    \begin{@twocolumnfalse}
      \begin{abstractbox}
        \maketitle
        \vspace{-1.1cm}
        \noindent{\sffamily\bfseries Abstract.}\enspace
        \abstracttext
      \end{abstractbox}
      \vspace{4mm}
    \end{@twocolumnfalse}
  ]
  \makeatother
\fi

\input{body}

\section*{Data Availability}
The code used to set up, run, and analyze the simulations and the CG ligand structures and topologies are available at \href{https://github.com/BereauLab/martini-protein-ligand-binding-free-energies}{github.com/BereauLab/martini-protein-ligand-binding-free-energies}. One exemplary ligand simulation dataset per target is available at \href{https://dx.doi.org/10.5281/zenodo.22893438}{doi: 10.5281/zenodo.22893438}.

\section*{Acknowledgments}
We thank Luis Itza Vazquez-Salazar and Sander Hummerich for valuable feedback on the manuscript.
T.B. acknowledges support by the Deutsche Forschungsgemeinschaft (DFG, German Research Foundation) under Germany's Excellence Strategy EXC 2181/1--390900948 (the Heidelberg STRUCTURES Excellence Cluster).
The authors used Claude by \href{https://claude.com/product/overview}{Anthropic} (Opus 4.8 and Opus 5.0) for language editing and proofreading of the manuscript.

\section*{Author Contributions}
L.J.W.: conceptualization, methodology, software, validation, formal analysis, investigation, data curation, writing--original draft, writing--review and editing, visualization.
T.B.: conceptualization, resources, writing--review and editing, supervision, project administration, funding acquisition.

\balance
\bibliography{references}

\onecolumn
\appendix
\setcounter{section}{0}
\renewcommand{\thesection}{S\arabic{section}}
\renewcommand{\theHsection}{S\arabic{section}}
\setcounter{subsection}{0}
\renewcommand{\theHsubsection}{\theHsection.\arabic{subsection}}
\setcounter{figure}{0}
\renewcommand{\thefigure}{S\arabic{figure}}
\renewcommand{\theHfigure}{S\arabic{figure}}
\setcounter{table}{0}
\renewcommand{\thetable}{S\arabic{table}}
\renewcommand{\theHtable}{S\arabic{table}}
\setcounter{equation}{0}
\renewcommand{\theequation}{S\arabic{equation}}
\renewcommand{\theHequation}{S\arabic{equation}}
\phantomsection
\pdfbookmark[1]{Supporting Information}{si}
\hypersetup{bookmarksdepth=0}
\input{appendix}

\end{document}

%% file: body.tex
\section{Introduction}
\label{sec:introduction}

\begin{figure*}[ht]
    \centering
    \includegraphics[width=\linewidth]{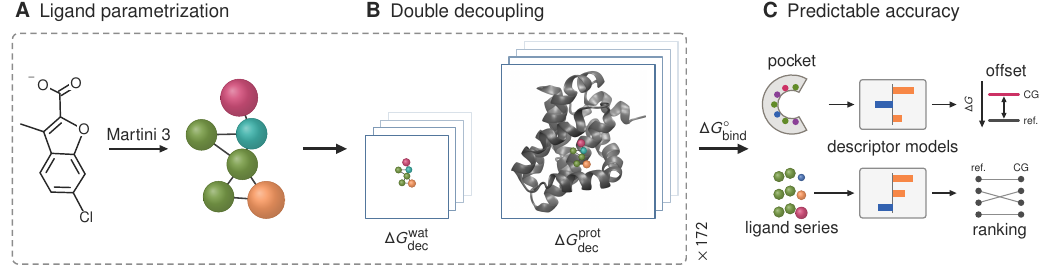}
    \caption{Overview of our workflow for computing and analyzing absolute binding free energies with the Martini~3 CG model.
        (A) Each of 172 ligands across 15 protein targets is coarse-grained into a Martini~3 model and (B) its binding free energy is computed by double decoupling in water and in the binding pocket (dashed box).
        (C) The systematic offset of the absolute free energies and the quality of the ligand ranking are then explained in terms of descriptors of the pocket and of the ligand series.}
    \label{fig:overview}
\end{figure*}

The binding free energy is a fundamental statistical-mechanics property characterizing the affinity between a ligand molecule and a protein.\cite{Gilson2007FreeEnergyReview}
It is directly related to the experimentally observable equilibrium binding constant through the Gibbs equilibrium relation.
While physics-based simulations can compute this free energy with close-to-experimental accuracy,\cite{Ross2023SchroedingerFEP} the cost of both accurate simulations and experimental measurements limits the number of molecules that can be assessed.
In the search for novel strong binders, it is therefore infeasible to cover the full chemical space of potentially relevant molecules.\cite{MedinaFranco2025BioReCS}
To make high-throughput screening practical, less expensive methods either rely on empirically fitted affinity scores\cite{Huang2010ScoringFunctions, Meli2022DLScoringFunctions} or trade physical rigor for speed.\cite{Genheden2015MMPBSA}
Their predictions are correspondingly less accurate,\cite{Su2018CASF2016} and the fitted scoring functions are often biased toward their training data.\cite{Sieg2019VSBiasControl}

Coarse-grained (CG) models offer a different compromise: they reduce the cost of simulation by orders of magnitude while still admitting a rigorous statistical-mechanical treatment.\cite{Noid2013CoarseGrainedModels}
The savings stem from a reduced representation in which groups of atoms are mapped onto effective interaction sites, or beads.
Fewer degrees of freedom mean fewer interactions to evaluate, while the smoother free-energy landscape permits larger time steps and accelerates the dynamics and sampling convergence.
For a given mapping, the effective potential that exactly reproduces the equilibrium distribution of the CG coordinates is the many-body potential of mean force (PMF), obtained by integrating out the atomistic degrees of freedom at fixed CG coordinates.\cite{Noid2008MultiscaleCoarseGraining}
Because the PMF is a high-dimensional, state-point-dependent free energy, it is intractable to compute exactly and must instead be approximated, typically by a sum of low-order terms.
The accuracy with which the resulting simulations reproduce equilibrium properties is therefore governed by the quality of this approximation.

The CG force field Martini approximates the PMF using simple bonded potentials, fitted to higher-resolution reference data, and non-bonded interactions between a fixed, discrete set of bead types.\cite{Souza2021Martini3}
The non-bonded interaction between any pair of bead types is identical across systems, making the force field transferable:
a molecule is parametrized once and can be reused without refitting.
Because many molecules map onto the same bead types, Martini also compresses chemical space, reducing the number of calculations needed to explore a given region.\cite{Menichetti2019ChemicalSpacePermeability, Kanekal2019ChemicalSpaceResolution}
The parameters of the non-bonded interactions are calibrated against partitioning free energies between phases of different polarity.\cite{Souza2021Martini3}
While this top-down parametrization inevitably loses some chemical detail---motivating several specialized, less transferable Martini variants\cite{Thomasen2024RescalingProteinProteinInteractions, ClaverasCabezudo2023ScalingProteinWaterInteractions, Lunt2025MartiniPEI}---its standard form still retains much of the physics underlying biological processes.
Martini has correspondingly been applied to a wide range of biomolecular systems, from lipid membranes\cite{Pedersen2025MartiniLipidome} and proteins\cite{BorgesAraujo2024AssessingMartini3ProteinModel} to carbohydrates.\cite{Grunewald2022Martini3Carbohydrates}

Motivated by its partitioning-based calibration, Martini has repeatedly been suggested for calculating protein--ligand binding affinities.\cite{Souza2021LigandProteinPerspectives, Kjolbye2022TowardsDesignDrugs, Marrink2023TwoDecadesMartini, BorgesAraujo2024AssessingMartini3ProteinModel}
While multiple studies address protein--ligand binding,\cite{Madaloz2026MartiniSteroidHormones, Dandekar2020ProteinLigandRecognition, Waclawikova2023GutBacterialMetabolite} only a few consider binding free energies, and these are limited to a small number of targets and ligands.\cite{Souza2020ProteinLigandBinding, Grazzi2026FunnelMetadynamics1, Grazzi2026FunnelMetadynamics2, Bartocci2024MartiniCannabinoidBinding}
For example, using PMFs from unbiased simulations, \citeauthor{Souza2020ProteinLigandBinding} reproduced experimental absolute binding free energies for seven ligands to T4 lysozyme mutants with a mean absolute error (MAE) of \SI{0.3}{\kilo\calorie\per\mole}.\cite{Souza2020ProteinLigandBinding}
More recently, \citeauthor{Grazzi2026FunnelMetadynamics1} used funnel metadynamics to compute binding free energies for selected protein--ligand pairs from two targets and several ligands, reporting a \SI{2}{\kilo\calorie\per\mole} MAE against their own atomistic simulations.\cite{Grazzi2026FunnelMetadynamics1}
However, Martini's accuracy on binding free energies has not been assessed systematically across a broad range of targets and ligands, leaving it unclear how well its partitioning-based parametrization transfers to protein--ligand binding.

In this work, we compute absolute binding free energies with Martini~3 for 172 ligands across 15 protein targets, using validated CG ligands.
The free energies are computed with a sampling-efficient alchemical decoupling protocol adapted for the faster dynamics and the less sharply defined bound state of CG models.
As in high-throughput screening, we use a single, largely automated workflow for the CG parametrization and the simulation setup, which also makes our results independent of system-specific fine-tuning.
We validate the protocol against CG free energies from unbiased simulations.
As a reference, we use the datasets of \citeauthor{Ross2023SchroedingerFEP} and \citeauthor{Baumann2026OpenFE}, which provide accurate experimental and atomistic binding free energies.\cite{Ross2023SchroedingerFEP, Baumann2026OpenFE}

We show that the accuracy of the Martini binding free energies, both of the absolute values and of the ligand ranking, depends on the target, the ligand, and the physical factors driving their interaction.
We find that Martini systematically underbinds across all tested targets, with the degree of underbinding being predictable from a simple linear model in the number of charged pocket residues.
For the ligand ranking, the accuracy can be anticipated from the structure--activity relationship driving the affinity variation within a series: hydrophobicity and ligand size are captured accurately, while polar interactions are less well represented.
Compared to other approximate affinity prediction methods such as molecular mechanics generalized Born surface area (MMGBSA), Martini's coarse-graining removes some of the chemical detail needed to rank similar ligands, but retains the statistical-mechanical basis required for physically meaningful free energies.
Overall, the physics-based parametrization of Martini allows us to interpret and predict the accuracy of the computed binding free energies based on the underlying mechanisms of binding between the protein and the ligand.

\Cref{sec:methods:abfe} outlines our general approach to computing absolute binding free energies, including considerations specific to the CG Martini model.
The rest of \cref{sec:methods} then covers the reference dataset, ligand parametrization, and remaining technical details of the protocol, with \cref{fig:overview} summarizing the full workflow.
We present the resulting binding free energies and discuss their implications in \cref{sec:results}.

\section{Methods}
\label{sec:methods}

\subsection{Coarse-Grained Absolute Binding Free Energies}
\label{sec:methods:abfe}
Since coarse-graining approximates the interactions but leaves the statistical mechanics unchanged, the same free-energy methods used for atomistic simulations remain applicable.
While Martini binding free energies have so far been computed mainly with pathway methods,\cite{Chipot2023PathwayMethods,Souza2020ProteinLigandBinding,Grazzi2026FunnelMetadynamics1} we used alchemical methods, which connect the bound and unbound states through non-physical intermediates, for their simpler setup and higher sampling efficiency.\cite{Mey2020BestPracticesAlchemical,Ruiter2011ProteinLigandFreeEnergies}

\begin{figure}[t]
    \centering
    \includegraphics[width=\linewidth]{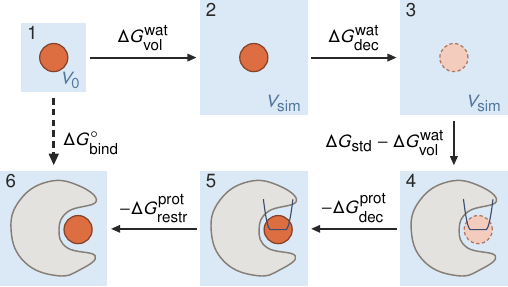}
    \caption{Alchemical thermodynamic cycle for absolute protein--ligand binding free energies.
        The standard binding free energy $\Delta G^{\circ}_{\mathrm{bind}}$ (dashed arrow) is obtained by summing the legs $1\rightarrow6$.
        The ligand (orange circle; faded when decoupled) is alchemically decoupled from solvent in the water leg ($2\rightarrow3$, $\Delta G_{\mathrm{dec}}^{\mathrm{wat}}$) and recoupled in the binding site in the protein leg ($4\rightarrow5$, $-\Delta G_{\mathrm{dec}}^{\mathrm{prot}}$), where a flat-bottom restraint (well) keeps the ligand center of mass in the binding site in both states.
        The solvent-volume term $\Delta G_{\mathrm{vol}}^{\mathrm{wat}}$ ($1\rightarrow2$, $V_0\rightarrow V_{\mathrm{sim}}$) cancels against step $3\rightarrow4$;
        the standard-state correction $\Delta G_{\mathrm{std}}$ is evaluated analytically and the restraint term $\Delta G_{\mathrm{restr}}^{\mathrm{prot}}$ ($5\rightarrow6$) is obtained by reweighting to the bound state defined by the flat-bottom radius.}
    \label{fig:thermodynamicCycle}
\end{figure}

Specifically, we used the double-decoupling method,\cite{Gilson1997DoubleDecoupling} in which the ligand's non-bonded interactions are turned off separately in the bound ($\Delta G_\mathrm{dec}^\mathrm{prot}$) and unbound ($\Delta G_\mathrm{dec}^\mathrm{wat}$) states.
Each decoupling was discretized into intermediate states along a coupling parameter $\lambda$.\cite{Mey2020BestPracticesAlchemical}
In the bound leg, turning off these interactions also removes the forces that hold the ligand in the binding pocket.
A restraint was therefore applied to keep the ligand localized in the binding site throughout the decoupling.\cite{Deng2009BindingFreeEnergies}
Since it also acts in the fully interacting end state, the restraint must either leave the bound ensemble unbiased or be removed explicitly ($\Delta G_\mathrm{restr}^\mathrm{prot}$).
The resulting free energy of the bound leg depends on the ligand's accessible volume, and hence on the shape of the restraining potential.
An analytical correction $\Delta G_\mathrm{std}$ removes this dependence by referencing the free energy to the standard-state concentration of \SI{1}{\molar}.
For the unbound leg, the corresponding term $\Delta G_\mathrm{vol}^\mathrm{wat}$ cancels from the cycle, since both end states sample the same volume.
The complete thermodynamic cycle is shown in \cref{fig:thermodynamicCycle}, and yields
\begin{equation}
    \Delta G_\mathrm{bind}^{\circ} = \Delta G_\mathrm{dec}^\mathrm{wat} - \Delta G_\mathrm{dec}^\mathrm{prot} - \Delta G_\mathrm{restr}^\mathrm{prot} + \Delta G_\mathrm{std}.
    \label{eq:abfe}
\end{equation}

While the general protocol is equivalent to the atomistic case, the choice of restraint requires particular care in the CG setting.
Atomistic protocols typically use Boresch-style restraints, which fix the ligand's position and orientation relative to the protein.\cite{Boresch2003Restraints}
Since CG models sample the pocket faster, this orientational restriction is not needed for a converged bound leg.
In addition, such a restraint requires an accurate binding pose for each complex.
We therefore restrained only the distance between the center of mass of the ligand and the binding site.

The same accelerated dynamics that improve sampling efficiency require the restraint to remain active even in the fully interacting state.
In atomistic simulations, the restraint is often switched off in this state, relying on kinetic trapping of the ligand in the pocket.\cite{Mey2020BestPracticesAlchemical}
In Martini, this assumption breaks down: the fast dynamics, combined with a general tendency of CG ligands to underbind (\cref{sec:results:targetDependence}), substantially increase the ligand's chance of leaving the pocket over the course of the simulation, even in the fully interacting state.
We therefore kept the restraint active in every window.
The bound state is then defined not by the ligand's interactions with the protein alone but also by the restraint, which must therefore be chosen carefully to avoid biasing the sampled configurations.
We used a flat-bottom potential that exerts no force within the pocket and confines the ligand only near the edge of the bound region.
Given sufficient sampling within the flat-bottom region, the samples can be reweighted to the bound state defined by the flat-bottom radius, discarding all configurations sampled beyond it.\cite{Rizzi2020Reweighing}
This reweighting yields the restraint correction $\Delta G_\mathrm{restr}^\mathrm{prot}$.
Substantial sampling against the restraint wall signals an inadequate restraint setup.
Further details on the restraint and reweighting procedure are provided in \cref{si:methods:restraints} of the Supporting Information.

\subsection{Reference Dataset and System Selection}
\label{sec:methods:referenceDataset}
For our assessment of Martini binding free energies, we relied on the established dataset of \citeauthor{Ross2023SchroedingerFEP} as our main reference, thereby avoiding biases from self-run atomistic reference calculations.\cite{Ross2023SchroedingerFEP}
They collected experimental binding free energies with at least two measurements each for a large set of protein--ligand complexes and compared them against carefully prepared atomistic relative binding free energies from the FEP+ program.
Their reported pairwise root-mean-square error (RMSE) of \SI{1.25}{\kilo\calorie\per\mole} relative to the experimental values is only slightly larger than the pairwise RMSE of \SI{0.91}{\kilo\calorie\per\mole} between repeated experimental measurements, which effectively provides a lower bound on the achievable accuracy.
Because they achieved such good agreement with experiment, comparing our results to either reference yields similar conclusions.
Comparing with the atomistic reference, however, allows us to separate errors arising from the coarse-graining from those due to experimental uncertainty.
Since \citeauthor{Ross2023SchroedingerFEP} only computed relative binding free energies, these were anchored to an absolute scale by matching the per-target mean to the experimental values.
The atomistic reference therefore isolates CG errors within a target, such as in the ligand ranking, whereas the absolute offset of each target is effectively measured against experiment.

We selected a subset of 15 protein targets with up to 12 ligands each to cover a broad range of systems while limiting the ligand parametrization effort, the main bottleneck in our workflow.
We drew the targets from the \emph{fragments},\cite{Steinbrecher2015FragmentBFE} \emph{jacs},\cite{Wang2015JacsBFE} and \emph{merck}\cite{Schindler2020MerckBFE} subsets, deliberately excluding more challenging systems such as GPCRs, macrocyclic ligands, and systems with cofactors.
Where more than 12 ligands were available, 12 were selected randomly; two targets had only 11 ligands and one had only 6, giving 172 ligands across 15 targets in total.
Five targets included charged ligands, two of them a mix of charged and neutral, amounting to 48 charged and 124 neutral ligands.
The ligands span 6 to 38 heavy atoms (median 26).
For all targets, most ligands form congeneric series: they share a common scaffold and differ only in their substituents (\cref{si:ligandTopologies} of the Supporting Information).
This explains the often small dynamic range of binding free energies within a target, with per-target ranges spanning \SI{2.4}{\kilo\calorie\per\mole} to \SI{7.9}{\kilo\calorie\per\mole} (median \SI{3.6}{\kilo\calorie\per\mole}).
\Cref{si:methods:proteinTargets} of the Supporting Information gives a detailed overview of the selected targets and ligands.

Additionally, we used the dataset of \citeauthor{Baumann2026OpenFE}, who computed binding free energies for the same systems with the open-source program OpenFE.\cite{OpenFE2025, Baumann2026OpenFE}
We took their prepared atomistic protein and ligand structures, including binding poses and protonation states, as the starting point for our Martini simulations.\cite{Baumann2026OpenFE}
They also reported MMGBSA binding free energies computed with Prime,\cite{SchroedingerPrime} which we used as an additional reference for comparison.

\subsection{Ligand Parametrization}
\label{sec:methods:ligands}
The CG ligand models were built following the established protocols for Martini~3 small molecules.\cite{Souza2021Martini3, Alessandri2022MartiniSmallMolecules}
Given the 172 ligands in this study, we prioritized a protocol that reliably yields stable simulations for the whole set over fine-tuning individual ligands;
the accuracy of the resulting models is assessed in \cref{sec:results:ligandValidation}.
We did not use automated parametrization tools such as Auto-Martini,\cite{Magdalena2025AutoMartini3} given their limitations in molecule size and bead-type assignment.
For 7 ligands of the T4 lysozyme target (\emph{t4lys}), we used the parametrizations already available from previous work.\cite{Alessandri2022MartiniSmallMolecules}
Within each target, ligands were parametrized consistently, representing their common scaffold with identical bead types and bonded topologies.
For the atom-to-bead mapping we focused on preserving the symmetry of the atomistic structure.
We used mostly tiny (\texttt{T}) bead types, for high chemical detail and because the ligands are rich in aromatic rings.
Small (\texttt{S}) bead types were used for aliphatic rings and small branched substituents, and regular (\texttt{R}) bead types for larger, heavily branched substituents of 4 or 5 heavy atoms.
Bead types were assigned primarily from existing Martini~3 parametrizations of small molecules.\cite{Alessandri2022MartiniSmallMolecules, Grunewald2022Martini3Carbohydrates, Martini3Bible}
Where no reference was available, we selected the most appropriate bead type from the general selection rules,\cite{Souza2021Martini3} predicted partitioning free energies,\cite{Tetko2002ALOGPS} and the fragment's chemical environment.
For charged ligands, we additionally parametrized a neutral version, identical except for the bead type of the charged group, since reference partitioning data are available only for neutral compounds.
The CG mapping and bead-type assignment were performed with a custom tool inspired by CGBuilder.\cite{CGBuilder}
The CG topologies are visualized in \cref{si:ligandTopologies} of the Supporting Information.

Bonded parameters were derived by Boltzmann inversion of atomistic reference simulations (\cref{si:methods:atomisticReferenceSimulations}) using the program Fast-Forward.\cite{Brasnett2026FastForward, FastForward}
Constraints were assigned to aromatic rings and virtual sites to fused ring systems, as established for Martini~3 small molecules.\cite{Alessandri2022MartiniSmallMolecules}
To avoid large constraint networks that destabilize simulations, all remaining bonds were treated as harmonic, with force constants fitted to the reference simulations.

Angles and dihedrals were assigned for all remaining degrees of freedom, taking equilibrium values from the reference distributions and omitting redundant terms, which would otherwise over-stiffen the molecule under Boltzmann inversion.
We slightly modified Fast-Forward to improve simulation stability, replacing near-collinear harmonic angles with linear angle potentials\cite{vanDerSpoel2020LinearMoieties} and the proper and improper dihedral terms with a combined bending-torsion potential.\cite{Bulacu2005BendingTorsionRigidity, Bulacu2013ImprovedAnglePotentials}
Further details are given in \cref{si:methods:ligandParametrization} of the Supporting Information.

We validated the ligand models by comparing their solvent-accessible surface area (SASA), octanol--water partitioning, and bonded distributions with atomistic references.
The SASA values were computed using the GROMACS tool \texttt{gmx sasa}, with atomistic radii from \citeauthor{Rowland1996AtomRadii}.\cite{Rowland1996AtomRadii,Alessandri2022MartiniSmallMolecules}
Because experimental octanol--water partitioning free energies were not available for most ligands, we used converted logP predictions from ALOGPS,\cite{Tetko2002ALOGPS} XLOGP3,\cite{Cheng2007XLOGP3} MLOGP,\cite{Moriguchi1992MLOGP, Lipinski2001MLOGP} and SILICOS-IT,\cite{SilicosIT} the last three obtained from the SwissADME web service.\cite{Daina2017SwissADME}
Distributions of bonds, angles, and dihedrals between CG and atomistic simulations were compared using the \texttt{ff\_assess} scoring method by \citeauthor{Brasnett2026FastForward}, which combines the Hellinger distance with a penalty for deviations from the distribution mean.\cite{Brasnett2026FastForward}
Further details on the three metrics are given in \cref{si:methods:validation} of the Supporting Information.

\subsection{Simulation Details}
\label{sec:methods:simulationProtocol}
Protein structures and their protonation states were taken from the atomistic reference structures prepared by \citeauthor{Baumann2026OpenFE}\cite{Baumann2026OpenFE}
They were coarse-grained and parametrized with Martinize2,\cite{Kroon2024Martinize2} using the OLIVES G\={o}-model\cite{Pedersen2024OLIVES} for structural bias;
OLIVES has been reported to reproduce protein--ligand binding better than alternative biasing schemes.\cite{Madaloz2026MartiniSteroidHormones,Grazzi2026FunnelMetadynamics2}
Martinize2 assigned charges from the input protonation states, which were thus retained in the CG models.
Simulation systems were solvated in a cubic box using packmol.\cite{Martinez2009Packmol}

To restrain the ligand center of mass within the binding pocket, we applied a flat-bottom restraint with a radius of \SI{0.4}{\nano\meter} and a force constant of \SI{2000}{\kilo\joule\per\mole\per\nano\meter\squared}.
The binding pocket was defined by a non-interacting dummy bead placed at the ligand center of mass in the bound pose from the atomistic reference structure.
The bead was constructed as a virtual site from four to eight backbone beads of the protein residues lining the pocket.
The comparatively small radius is appropriate since the restraint acts only on the center of mass, not the full extent of the ligand, and the ligand already started in a reasonable bound pose (see \cref{si:results:flatBottomRadius} of the Supporting Information).

For the ligand decoupling, we used 20 and 17 non-linearly spaced $\lambda$ windows to decouple the van der Waals interactions in the protein and water legs, respectively.
For charged ligands, three or two additional $\lambda$ windows were used to decouple the electrostatic interactions in the protein and water legs, respectively.
The $\lambda$ values were chosen to ensure similar overlap between adjacent windows (see \cref{si:methods:lambdaSchedule} of the Supporting Information for further details).
We used Hamiltonian replica exchange to enhance sampling, with exchanges between adjacent windows attempted every \SI{20}{\pico\second}.
This reduces kinetic trapping in metastable states, which can occur despite the fast dynamics of the Martini~3 CG model.

All simulations used the Martini~3 force field\cite{Souza2021Martini3} with GROMACS 2026.0\cite{Abraham2025GROMACS} and standard Martini simulation parameters.
Electrostatics were treated with the Reaction-Field method ($\varepsilon_{\mathrm{rf}}=0$),\cite{Tironi1995GeneralizedReactionField} which simplifies the decoupling of charged ligands by removing the need for a reverse-coupled neutralizing counterion.
We observe no charge-specific deviation from the reference (\cref{si:results:offsetChargeTest} of the Supporting Information).
Further details on the simulation protocol and restraint setup are provided in \cref{si:methods:restraints,si:methods:simulationDetails} of the Supporting Information.

\subsection{Free-Energy Estimation}
\label{sec:methods:freeEnergyEstimation}
We estimated free energies from the simulation data using the multistate Bennett acceptance ratio (MBAR),\cite{Shirts2008MBAR} as implemented in PyMBAR.\cite{Wu2024Alchemlyb, Shirts2025pymbar}
We discarded the first \SI{5}{\percent} of each simulation as equilibration and kept only decorrelated frames from the remainder.
For the protein leg we estimated the decorrelation time from the distance to the pocket center, since the reduced potential decorrelates too quickly to capture the slow pocket dynamics.
For the faster-relaxing water leg we used the reduced potential.
Because the MBAR uncertainties were still consistently smaller than the spread across independent repeats, we reported the standard error of the mean across repeats as the final uncertainty: three repeats for the protein leg and two for the water leg.

\subsection{Pocket and Ligand Descriptors}
\label{sec:methods:descriptors}
To understand how Martini's protein--ligand binding free energies depend on the individual targets and ligand series, we characterized both by a set of descriptors that classify interactions by their driving mechanism, such as hydrophobicity, hydrogen bonding, or $\pi$-stacking.\cite{Bissantz2010MedicinalChemistsGuide}
We thereby aimed to identify which of these driving forces Martini captures well and which it does not, relating this back to Martini's parametrization approach.
We used two correlation metrics to quantify the quality of the ranking and relate the descriptors to the binding free-energy results:
the Spearman correlation $\rho$ measures the monotonicity of a relationship, i.e., whether the order of the values is preserved, while the Pearson correlation $r$ describes the linearity of a relationship and yields the coefficient of determination $R^2 = r^2$.
Error bars for correlation coefficients are $\pm 1$ standard error computed on the Fisher $z$-scale.\cite{Fisher1921ProbableErrorCorrelation, Fieller1957RankCorrelationCoefficients, Bonett2000SampleSizeRequirements}

Computing pocket descriptors for a protein target requires a defined binding pocket, which we derived from the bound ligands by considering all residues with a heavy atom within \SI{0.6}{\nano\meter} of any ligand heavy atom as pocket residues.
This cutoff covers the typical range of all non-covalent interactions.\cite{Bissantz2010MedicinalChemistsGuide}
We reported per-target averages over all ligands and used the variation over the ligand set as error bars.
We considered five descriptors: the number of pocket residues, i.e., the pocket size, and the fractions of charged, aromatic, polar, and non-polar residues, obtained by dividing the respective residue counts by the pocket size.
Residue classes are not mutually exclusive, e.g., tyrosine counts as aromatic and polar.
Together with the normalization of the fractions by the pocket size, this renders the descriptors partially collinear.
Charged residues were assigned from the protonation states of the atomistic structures prepared by \citeauthor{Baumann2026OpenFE}\cite{Baumann2026OpenFE}
See \cref{si:methods:pocketAndLigandDescriptors} of the Supporting Information for the composition of each residue class.
Taken together, the descriptors serve as proxies for the pocket's capacity for the different binding mechanisms.

We characterized each ligand series by its structure--activity relationship (SAR), i.e., by how strongly the binding free energy varies with a given ligand feature, quantified as the Spearman correlation between the reference binding free energies and that feature.
The ligand features comprise intrinsic properties---the predicted logP, the number of heavy atoms, and the charge---and pocket-dependent interactions such as hydrogen bonding and $\pi$-stacking.
Other features we tested varied across too few series or did not show a significant correlation with the ranking accuracy (\cref{si:results:structureActivityRelationships} of the Supporting Information).
The pocket-dependent interactions were computed with PLIP\cite{Salentin2015PLIP} on the atomistic input structures.

\section{Results and Discussion}
\label{sec:results}

\subsection{Validation of the Ligand Parametrization}
\label{sec:results:ligandValidation}
First, we validate our CG ligand models.
All ligands pa\-ram\-e\-trized for this study simulate stably at the standard Martini time step of \SI{20}{\femto\second}.
The following validation results provide only an overview, with further details given in \cref{si:results:ligandValidation} of the Supporting Information.

We compare the octanol--water partitioning free energies of the neutral CG ligands against the reference from the four logP prediction models (\cref{sec:methods:ligands}).
The CG ligands reproduce these values with an MAE of \SI{0.6}{\kilo\calorie\per\mole} and a median per-target Spearman correlation of \SI{0.81}{}.
The prediction methods themselves agree only to a median per-target pairwise Spearman correlation of \SI{0.88}{}, which bounds the agreement attainable against such an approximate reference.

As a second metric, we compare the SASA of the CG ligands against the atomistic reference.
Here too the deviations are small, with a median relative error of \SI{2.4}{\percent}.
Overall, the CG SASA values are lower than the atomistic reference, giving a mean signed relative error of \SI{-1.6}{\percent}.
This likely results from focusing on tiny and small bead types to preserve chemical detail, which reduces the effective SASA per atom compared to regular-sized beads (\cref{si:results:ligandValidation}).
Although increasing bond lengths or bead sizes could improve the SASA agreement, this would require modifying equilibrium parameters derived from atomistic simulations or bead radii fixed by the Martini force field.

Finally, we compare the bonded distributions based on the \texttt{ff\_assess} score.
All but two ligands score below \SI{0.5}{}, indicating good or acceptable agreement with the atomistic reference (\cref{fig:si:ligandValidation,fig:si:bondedScoreDistribution}).
Higher scores arise mainly from angle distributions that have a median score of \SI{0.43}{}, despite the Fast-Forward-based Boltzmann-inversion parametrization.
Importantly, neither SASA errors nor low bonded-distribution overlap correlates with binding free-energy errors (\cref{fig:si:residualVsLigandMetrics}).
Although we did not iteratively fine-tune the CG parametrizations, these results show that the ligand models reproduce the reference properties well and are therefore suitable for further simulations.

\subsection{Validation of the Free-Energy Protocol}
\label{sec:results:freeEnergyValidation}
Next, we validate our double-decoupling free-energy protocol.
Assessing the accuracy of Martini binding free energies relative to the atomistic reference requires separating errors arising from the model from those introduced by the simulation protocol.
First, we compare our protocol against the T4 lysozyme results of \citeauthor{Souza2020ProteinLigandBinding}, who computed absolute binding free energies for seven small ligands using unbiased simulations.\cite{Souza2020ProteinLigandBinding}
Because they used the open-beta version of Martini~3, which differs substantially from the official release, we used those force field parameters to recompute the same systems with our protocol.
Our results agree well with the reference, with an MAE of \SI{0.52}{\kilo\calorie\per\mole} (see \cref{si:results:souzaComparison} of the Supporting Information for details), despite running only \SI{1.46}{\micro\second} of simulation per ligand compared to the \SI{900}{\micro\second} of the reference simulations.

As a second validation, we ran unbiased simulations with the official Martini~3 parameters for five ligands across four targets to compare them against our double-decoupling results.
To improve sampling efficiency, we restricted the accessible volume to a sphere of \SI{3.5}{\nano\meter} radius around the binding pocket.
We also focused on relatively weak binders, since stronger binders would require much longer simulations to observe unbinding events.
These choices yielded reasonably converged free energies within \SI{3}{\micro\second} (for the \emph{t4lys} system) to \SI{9}{\micro\second} (for the other systems) of simulation time per ligand.
The unbiased free energies agree well with those from our double-decoupling protocol, with a mean signed error of \SI{-0.07}{\kilo\calorie\per\mole} and an MAE of \SI{0.30}{\kilo\calorie\per\mole} (\cref{si:results:unbiasedSimulationResults}).

To assess the convergence of our double-decoupling protocol, we ran three independent repeats for each ligand in the protein leg and two repeats in the water leg.
Across all 172 ligands, the repeats agree to a median standard error of \SI{0.18}{\kilo\calorie\per\mole} in the protein leg and \SI{0.06}{\kilo\calorie\per\mole} in the water leg, giving a median uncertainty of \SI{0.23}{\kilo\calorie\per\mole} for the binding free energy.
For five neutral ligands from five targets, we recomputed the free energies with coarser $\lambda$ schedules and from subsets of the trajectories.
Halving the number of windows changes the free energies by at most \SI{0.06}{\kilo\calorie\per\mole} in either leg, indicating sufficient overlap between adjacent windows.
Reducing the sampling time has a slightly larger effect in the protein than in the water leg:
retaining only the last \SI{60}{\percent} of each trajectory changes the free energy by \SI{0.27}{\kilo\calorie\per\mole} in the protein leg but only \SI{0.06}{\kilo\calorie\per\mole} in the water leg, despite the latter being sampled for only \SI{2.4}{\nano\second} per window compared to \SI{50}{\nano\second} in the protein leg (\cref{si:results:convergence}).
Although this \SI{0.27}{\kilo\calorie\per\mole} shift suggests the protein leg is reasonably converged, the differences between repeats are sometimes larger, with a 90th percentile of the standard error of \SI{0.41}{\kilo\calorie\per\mole}.
This indicates that ligands occasionally sample different dominant bound poses.
Instead of extending the protein simulations, running multiple repeats captures the variation more efficiently and provides a more reliable uncertainty estimate than a single long trajectory.

This pose variation likely reflects a general property of the CG model, whose coarse and isotropic interactions define the bound state less sharply than the atomistic reference.
Even with our narrow flat-bottom pocket restraint ($r_{\mathrm{restr}}=\SI{0.4}{\nano\meter}$), independent repeats can settle on slightly different dominant poses.
Although increasing the restraint radius lowers the free energy due to the larger accessible volume, it also amplifies this variation and allows sampling of states that are not representative of the native bound pose (\cref{si:results:flatBottomRadius} of the Supporting Information).
This could reduce the specificity of the binding free energies and thereby weaken the ranking of ligands.
We therefore kept the tight flat-bottom restraint, which confines the ligand to the native binding mode.

Beyond convergence, the computational cost required to compute a binding free energy is a further important consideration when comparing different methods.
Although simulation times depend on the required level of convergence and are therefore difficult to compare directly, they still provide a rough estimate of the computational cost.
For the binding free energies computed in this study, the cost per ligand ranged from \SI{2.2}{} GPU hours for \emph{mup1} to \SI{7.0}{} GPU hours for \emph{cdk8}, with a mean of \SI{3.8}{} GPU hours across all targets.
All replica-exchange simulations per ligand and leg were run on two NVIDIA A30 GPUs.
In contrast, the atomistic relative binding free-energy calculations of \citeauthor{Baumann2026OpenFE} required between \SI{8.7}{} and \SI{77.5}{} GPU hours per ligand, with a mean of \SI{28.8}{} GPU hours.\cite{Baumann2026OpenFE}
Because relative calculations require less sampling than absolute binding free energies and the reference calculations ran on faster GPUs, the observed speedup of about one order of magnitude is a lower bound (\cref{si:methods:runtimeComparison}).

\begin{figure}[t]
    \centering
    \includegraphics[width=\linewidth]{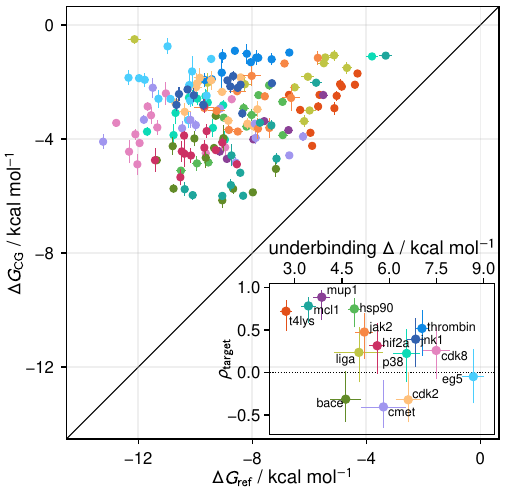}
    \caption{Absolute binding free energies $\Delta G_{\mathrm{CG}}$ computed with Martini~3 for 172 ligands across 15 targets, compared against the FEP+ reference values $\Delta G_{\mathrm{ref}}$. All values lie above the diagonal, indicating that Martini predicts weaker binding than the reference. The inset shows the mean underbinding $\Delta$, i.e., the offset from the reference, and the ligand ranking Spearman correlation $\rho_{\mathrm{target}}$ for each target. All error bars indicate one standard error.}
    \label{fig:targetDependentAccuracy}
\end{figure}

\subsection{Target-Dependent Binding Free-Energy Accuracy}
\label{sec:results:targetDependence}
With the protocol validated, we now assess the accuracy of the predicted binding free energies against the atomistic reference.
\Cref{fig:targetDependentAccuracy} presents the binding free-energy results $\Delta G_{\mathrm{CG}}$ for all 172 CG ligands across the 15 targets, compared against the FEP+ reference values $\Delta G_{\mathrm{ref}}$.
The calculated binding free energies are uniformly negative and lie in a range typical of protein--ligand affinities, as expected for favorable binding of the ligands.
However, Martini underestimates the binding affinities across all ligands and targets, with a mean offset from the reference, or underbinding $\Delta$, of \SI{5.7}{\kilo\calorie\per\mole}.
The underbinding varies between targets, from \SI{2.7}{\kilo\calorie\per\mole} to \SI{8.7}{\kilo\calorie\per\mole} for \emph{t4lys} and \emph{eg5}, respectively.
The deviation is therefore almost entirely a target-dependent shift: removing it reduces the MAE from \SI{5.7}{} to \SI{1.0}{\kilo\calorie\per\mole} (\cref{fig:si:freeEnergyStats}D).
The results of the individual ligands and targets are shown in \cref{si:results:detailedBindingFreeEnergies} of the Supporting Information.
We do not observe this underbinding for the \emph{t4lys} simulations with the open-beta version of Martini~3 (mean signed error of \SI{-0.10}{\kilo\calorie\per\mole} against the reference), which we ran only for \emph{t4lys} to compare our protocol against results from \citeauthor{Souza2020ProteinLigandBinding} (\cref{si:results:souzaComparison} of the Supporting Information).
With the official Martini parameters and the identical remaining setup, the same seven systems underbind by \SI{2.35}{\kilo\calorie\per\mole}.
This indicates that the observed underbinding is a property of the Martini~3 interaction parameters rather than of the CG modeling approach.
Because Martini~3 was parametrized as a compromise across many systems, a calibration targeted at protein--ligand binding could reduce the underbinding, at least for some targets.
Since the open-beta parameters have been superseded by the official Martini~3 release, we did not extend this comparison to the other targets.

To further analyze the accuracy of Martini's binding free energies, we compute the Spearman correlation $\rho_{\mathrm{target}}$ between the CG and reference binding free energies for each target.
Similar to the underestimation of absolute binding free energies, the ranking accuracy varies between targets, with $\rho_{\mathrm{target}}$ ranging from \SI{0.89}{} to \SI{-0.41}{} for \emph{mup1} and \emph{cmet}, respectively.
Five of the 15 targets reach $\rho_{\mathrm{target}} \geq 0.5$, while four show no predictive ranking ($\rho_{\mathrm{target}} < 0$).
Ranking these ligands is inherently difficult, since the ligands' binding free energies span a median dynamic range of only \SI{3.6}{\kilo\calorie\per\mole} (\cref{sec:methods:referenceDataset}).
This is especially true for an absolute binding free-energy method paired with a transferable CG model, which maps multiple chemical groups onto the same bead types.
The inset in \cref{fig:targetDependentAccuracy} shows the underbinding $\Delta$ and ranking accuracy $\rho_{\mathrm{target}}$ for each target.
Although targets with a high ranking accuracy tend to underbind less, the correlation between these two quantities is weak ($R^2 = 0.24$).
In the following, we further analyze the target-dependent binding free-energy accuracy to better understand which binding mechanisms Martini's partitioning-based parametrization reproduces.
If the deviations follow from the physics of the model, they can be anticipated for a new system from the mechanisms that drive its binding.

\begin{figure*}[ht]
    \centering
    \includegraphics[width=\linewidth]{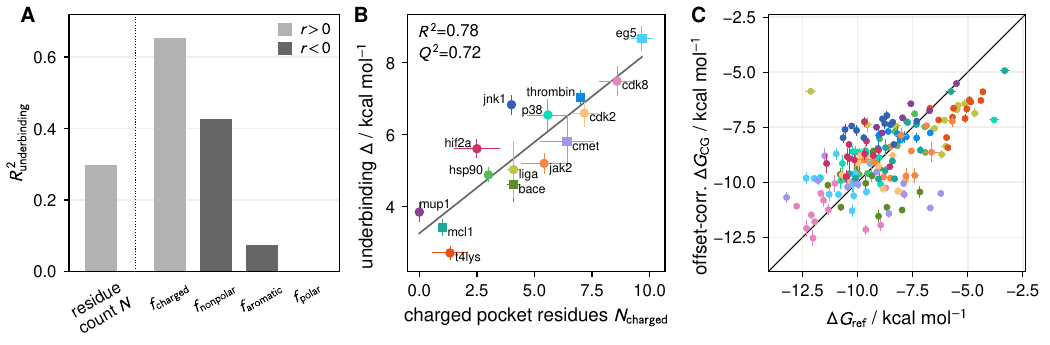}
    \caption{The number of charged pocket residues predicts the target-dependent underbinding $\Delta$. (A) Coefficient of determination $R^2$ between each pocket descriptor and $\Delta$; hatched bars mark negative correlations $r < 0$. (B) Underbinding $\Delta$ against the number of charged pocket residues, i.e., residue count times $f_{\mathrm{charged}}$, together with the linear model (line), its coefficient of determination $R^2$, and the leave-one-out cross-validated $Q^2$. (C) CG binding free energies corrected by the leave-one-out predictions of the model in (B), compared against the reference values $\Delta G_{\mathrm{ref}}$. Colors indicate the individual targets. All error bars indicate one standard error.}
    \label{fig:offsetModel}
\end{figure*}

\subsection{Pocket Properties Predict Underbinding}
\label{sec:results:pocketProperties}
First, we focus on the observed underbinding and the question of why it varies between targets.
The target dependence suggests that underbinding arises from the physicochemical properties of the pocket that determine ligand recognition.
We therefore characterize the 15 binding pockets by their size, i.e., the number of pocket residues $N$, and by the fractions of charged ($f_{\mathrm{charged}}$), non-polar ($f_{\mathrm{nonpolar}}$), aromatic ($f_{\mathrm{aromatic}}$), and polar ($f_{\mathrm{polar}}$) residues (\cref{sec:methods:descriptors}).
Strictly, the underbinding is a property of the protein--ligand complex rather than of the pocket alone.
However, within each target, the ligands bind with comparable, high affinity, so the dominant interaction pattern is set by the pocket.
Restricting the analysis to these target-dependent descriptors keeps the resulting insights independent of ligand-specific features.

We correlate each descriptor with the per-target underbinding $\Delta$ (\cref{fig:offsetModel}A).
The underbinding $\Delta$ correlates most strongly with $f_{\mathrm{charged}}$ ($R^2 = 0.65$), followed by $f_{\mathrm{nonpolar}}$ ($R^2 = 0.43$) and the pocket size $N$ ($R^2 = 0.30$).
The aromatic and polar fractions show essentially no correlation ($R^2 < 0.1$).
The fractions $f_{\mathrm{charged}}$ and $f_{\mathrm{nonpolar}}$ are themselves correlated ($R^2 = 0.46$) and capture a similar effect from opposite sides, since pockets with many charged residues necessarily contain fewer non-polar residues.
Combining $f_{\mathrm{charged}}$ and the pocket size $N$ into the absolute number of charged pocket residues $N_{\mathrm{charged}}$ yields an even stronger correlation ($R^2 = 0.78$, \cref{fig:offsetModel}B), robust under leave-one-out cross-validation ($Q^2 = 0.72$).
The corresponding linear fit provides a simple model for the underbinding
\begin{equation}
    \Delta = (0.5\,N_{\mathrm{charged}} + 3.3)\,\si{\kilo\calorie\per\mole}.
\end{equation}
The underbinding can be estimated for a new target without running any simulation, since the protein structure alone determines $N_{\mathrm{charged}}$.

Correcting the CG free energies of each target by its leave-one-out predicted offset yields the values shown in \cref{fig:offsetModel}C, with an MAE and RMSE against the reference of \SI{1.2}{} and \SI{1.6}{\kilo\calorie\per\mole}, respectively, close to the \SI{1.0}{} and \SI{1.3}{\kilo\calorie\per\mole} obtained with the exact per-target offsets.
For comparison, a null model that predicts every ligand by the mean reference free energy of the remaining targets reaches an MAE and RMSE of \SI{1.6}{} and \SI{2.0}{\kilo\calorie\per\mole}, respectively, while the OpenFE results agree with the FEP+ reference to within \SI{0.7}{} and \SI{1.0}{\kilo\calorie\per\mole}.
The corrected free energies therefore deviate about \SI{0.5}{\kilo\calorie\per\mole} more from the reference than the two atomistic methods do from each other.

Overall, Martini underbinds least for apolar pockets, where binding is driven mainly by hydrophobicity, and most for pockets rich in charged residues, where electrostatic interactions dominate.
Since the underbinding also occurs for neutral ligands, it likely arises from Martini's Lennard-Jones interactions between charged and neutral beads.
These interactions are not calibrated for the low-dielectric environment of a binding pocket, where electrostatic interactions are screened much less than in bulk solvent.\cite{Gilson2007FreeEnergyReview}
Martini applies the same implicit screening regardless of environment, which is known to underestimate the strength of polar and charged interactions in media of lower polarity.\cite{Yesylevskyy2010PolarizableWater}
The intercept shows that pockets without charged residues also underbind, which indicates that the bulk calibration of all bead interactions does not adequately represent the much denser environment of a binding pocket.
The remaining scatter around the linear model may, for example, result from pocket flexibility, which we described with the same structural bias for all targets instead of adapting it to each individual system.
Although hydrogen bonds formed by polar residues are also electrostatic interactions, the absence of a correlation between $f_{\mathrm{polar}}$ and the underbinding suggests that errors in hydrogen bonding do not accumulate into a systematic offset.
This may be because hydrogen bonds typically contribute little to the overall protein--ligand binding free energy, acting instead as a specificity filter,\cite{Bissantz2010MedicinalChemistsGuide} which is more relevant for the ligand ranking.

\subsection{Ligand Series Predicts Ranking Accuracy}
\label{sec:results:ligandSeries}

\begin{figure*}[ht]
    \centering
    \includegraphics[width=\linewidth]{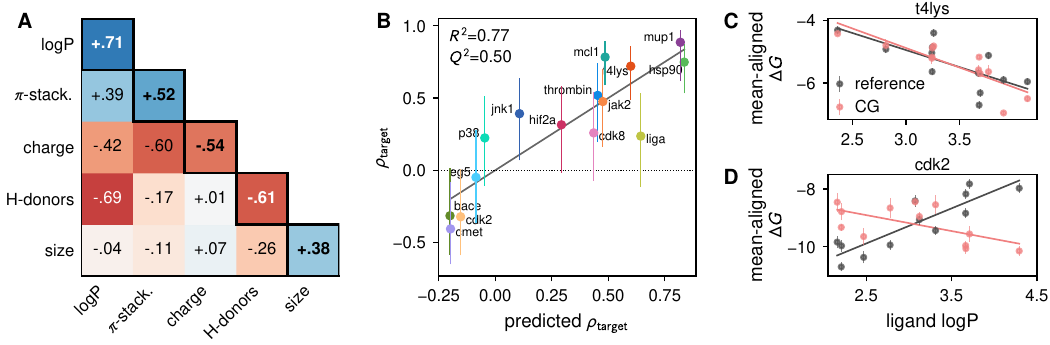}
    \caption{Structure--activity relationships (SARs) of a ligand series predict Martini's ranking accuracy $\rho_{\mathrm{target}}$. (A) Pearson correlation coefficients $r$ between the SARs. The diagonal entries give the correlation of each SAR with $\rho_{\mathrm{target}}$, the off-diagonal entries the cross-correlations between the SARs. (B) Ranking accuracy $\rho_{\mathrm{target}}$ observed for each target against the prediction of a linear model based on the logP, size, and $\pi$-stacking SARs. (C, D) Mean-aligned CG and reference binding free energies against the predicted ligand logP for \emph{t4lys} and \emph{cdk2} with linear fits.}
    \label{fig:rankingModel}
\end{figure*}

Next, we consider the ranking of the ligands within each series and the question of why it varies between targets.
Similar to the underbinding, the target dependence suggests that the ranking accuracy is determined by the physicochemical interactions between the ligands and the pocket rather than by the CG resolution alone.
To relate the ranking accuracy to physical binding mechanisms, we derive SARs $s_X$ for each series, i.e., the Spearman correlation between the reference binding free energies and a given ligand feature $X$ (\cref{sec:methods:descriptors}).
A large $|s_X|$ indicates that $X$ drives the affinity variation within that series, with the direction given by the sign of $s_X$.
Given the limited number of ligands per series, the SARs carry standard errors of \SI{0.3}{} (\SI{0.5}{} for \emph{mup1}), so we interpret them only qualitatively.

\Cref{fig:rankingModel}A shows the correlations between the SARs $s_X$ and Martini's ranking accuracy $\rho_{\mathrm{target}}$ on the diagonal, and the cross-correlations between the SARs themselves off the diagonal.
Martini generally ranks a series well when its affinity variation is driven by ligand hydrophobicity ($r = 0.71$), $\pi$-stacking ($r = 0.52$), or ligand size ($r = 0.38$).
It ranks poorly when the variation is driven by hydrogen-bond donors ($r = -0.61$) or ligand charge ($r = -0.54$).
We exclude the other SARs, such as the number of hydrogen-bond acceptors or salt-bridge interactions, which either vary across too few series or correlate only weakly with the ranking accuracy (\cref{fig:si:rankingHeatmapFull}).
The SARs are not independent of each other: the logP SAR correlates with those of the hydrogen-bond donors ($r = -0.69$), the charge ($r = -0.42$), and the $\pi$-stacking ($r = 0.39$).
Because logP aggregates the contributions of polar, charged, and aromatic groups, a series whose affinity varies with the number of hydrogen-bond donors will generally also vary with logP, only with opposite sign.

Combining the logP, size, and $\pi$-stacking SARs into a linear model reproduces the observed ranking correlation with $R^2 = 0.77$ (\cref{fig:rankingModel}B),
\begin{equation}
    \rho_{\mathrm{target}}=0.4\,s_{\mathrm{logP}} + 0.6\,s_{\mathrm{size}} + 0.5\,s_{\pi\text{-stack}} - 0.2.
\end{equation}
Despite the noise in the SARs, which limits the precision of the individual coefficients, the leave-one-out cross-validation ($Q^2 = 0.50$) indicates moderate transferability of the model to new targets.
Omitting any one of the three features reduces the model's $Q^2$ to \SI{0.39}{} or below, indicating that all three features contribute to predicting the ranking accuracy despite the limited number of targets.
The SAR-based model provides a qualitative, interpretable prediction of whether Martini can successfully rank a given ligand series.
A correction similar to that in \cref{sec:results:pocketProperties} is not possible for the ligand ranking, since ranking errors are not a uniform shift.
\Cref{fig:rankingModel}C and D illustrate the two limiting cases for \emph{t4lys} and \emph{cdk2}.
For \emph{t4lys}, both the reference affinities and the Martini values increase with the ligand logP, so the ranking is reproduced.
For \emph{cdk2}, Martini again predicts stronger binding for more hydrophobic ligands.
However, the reference affinities show the opposite trend, resulting in an inverted predicted ranking.
The SAR of \emph{cdk2} is dominated by hydrogen-bonding interactions ($s_{\text{H-donors}} = 0.82$), which the Martini model does not capture accurately.

These results are consistent with the pocket analysis above: Martini reproduces the hydrophobic and size contributions to binding, whereas it does not accurately model binding contributions from hydrogen bonding and charges.
For hydrogen bonds, this likely reflects the missing directionality of Martini's non-bonded interactions, which cannot distinguish a well-formed donor--acceptor geometry from a nonspecific contact and therefore blurs the small affinity differences within a series.
For charged groups, the bulk calibration of the interactions discussed above applies equally within a series.
Whether a ligand series can be ranked accurately therefore depends on how much of its affinity variation arises from hydrophobic and size effects.
Overall, Martini's physics-based parametrization makes it feasible to interpret and predict the accuracy of its binding free energies for a given target and ligand series.

\subsection{Comparison with MMGBSA}
\label{sec:results:comparison}
Finally, we compare Martini's binding free energies with those from MMGBSA.
Both methods can be used to replace expensive atomistic alchemical calculations, but differ substantially in their physical approximations.
MMGBSA combines an atomistic force-field energy with an implicit solvent model, treating the polar contribution with a generalized Born term and the nonpolar one through the SASA.\cite{Genheden2015MMPBSA}
The energy is evaluated only on structures of the bound and unbound states, instead of sampling the entire alchemical path between them.
This endpoint approach is orders of magnitude faster, but evaluates an energy instead of the ratio of configurational partition functions that defines the free energy.
Solvent entropy therefore enters only through the implicit solvation model, while the configurational entropy of protein and ligand is lost.
Since no configurational volume enters the calculation, the result cannot be referenced to a standard-state concentration.
MMGBSA nevertheless often ranks ligands within a series reasonably well.

\begin{figure}[t]
    \centering
    \includegraphics[width=\linewidth]{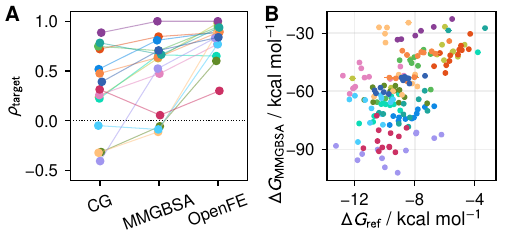}
    \caption{(A) Spearman correlation $\rho_{\mathrm{target}}$ of Martini, MMGBSA, and OpenFE against the FEP+ reference for each target.
        (B) MMGBSA absolute binding free energies $\Delta G_{\mathrm{MMGBSA}}$ computed by \citeauthor{Baumann2026OpenFE} against the reference values $\Delta G_{\mathrm{ref}}$. $\Delta G_{\mathrm{MMGBSA}}$ spans about eight times the range of the reference, with a mean absolute error of \SI{50.1}{\kilo\calorie\per\mole}.
        Note the different ranges of the two axes.
        Colors indicate the individual targets, as in \cref{fig:targetDependentAccuracy}.}
    \label{fig:spearmanRhosMMGBSA}
\end{figure}

Although the FEP+ free energies remain our main reference, we also include the OpenFE results, which provide a baseline for the variance between atomistic alchemical methods.
\Cref{fig:spearmanRhosMMGBSA}A compares the ranking accuracy $\rho_{\mathrm{target}}$ of Martini, MMGBSA, and OpenFE against the FEP+ reference for all 15 targets.
The OpenFE results correlate well with the reference for almost all targets (median $\rho_{\mathrm{target}} = \SI{0.89}{}$), while the MMGBSA and Martini results are more variable.
Overall, MMGBSA ranks ligands better than Martini, with a median $\rho_{\mathrm{target}}$ of \SI{0.64}{} versus \SI{0.32}{}.
Interestingly, the ranking accuracies of the two methods correlate across the 15 targets ($R^2 = 0.56$): both rank the same targets poorly, such as \emph{bace}, \emph{cdk2}, and \emph{eg5}.

\Cref{fig:spearmanRhosMMGBSA}B shows the MMGBSA absolute binding free energies against the reference values.
The mean absolute error of MMGBSA is \SI{50.1}{\kilo\calorie\per\mole}, substantially worse than Martini's \SI{5.7}{\kilo\calorie\per\mole} and OpenFE's \SI{0.7}{\kilo\calorie\per\mole}.
Much of this error stems from a systematic offset toward stronger binding, but even within a single target the spread of MMGBSA values remains too large to define a consistent scale (\cref{si:results:mmgbsaComparison} of the Supporting Information).
The systematic offset follows from the endpoint approach, which omits the unfavorable configurational entropy of binding and therefore overestimates the affinity.
Raw MMGBSA energies are thus not calibrated free energies and are used in practice only to rank ligands.

These comparisons show that MMGBSA and Martini occupy different regimes.
Although MMGBSA evaluates the full atomistic potential and preserves the fine chemical detail that separates closely related ligands, it does not yield a proper free energy.
Martini instead samples the full configurational ensemble, which yields physical binding free energies, while its CG interactions blur the chemical detail needed to rank congeneric ligands.
The coarse resolution is therefore a limitation when small structural differences must be resolved, but becomes an advantage when screening a large chemical space, where many atomistically distinct molecules map onto the same CG representation.
Even our congeneric ligand series, whose substituents were chosen to vary the affinity, show this effect: the 172 ligands map onto only 155 distinct combinations of bead types.
The \emph{t4lys} ligands m-xylene and o-xylene, for example, differ only in the geometry of otherwise identical CG models.
In general, the compression is much stronger: for molecules of up to two beads, \citeauthor{Menichetti2019ChemicalSpacePermeability} found this many-to-one mapping to reduce the number of required simulations by three to four orders of magnitude.\cite{Menichetti2019ChemicalSpacePermeability}
Since each bead type stands for several chemical fragments, the compression should grow further with the size of the molecules.
A single simulation then estimates the free energy of a whole neighborhood of ligands.
Although the required sampling makes a Martini binding free energy more expensive than an MMGBSA score, the GPU hours per ligand correspond to GPU seconds per represented molecule at a degeneracy of $10^3$.
More generally, Martini is only one possible resolution, and the balance between coverage and accuracy is set by how strongly a model compresses chemical space.\cite{Walter2025NavigatingChemicalSpace}

\section{Conclusion}
\label{sec:conclusion}
In this work, we study absolute binding free energies of 172 ligands across 15 targets computed with an alchemical double-decoupling protocol using the Martini~3 CG model.
We compare the results against atomistic alchemical free energies computed with FEP+ and OpenFE, as well as against MMGBSA.
We find that the accuracy of Martini's absolute binding free energies is systematically target- and ligand-dependent and predictable from the mechanisms that drive their binding.
Overall, Martini underestimates the binding affinities, with the underbinding being strongest for pockets rich in charged residues.
We derive a simple linear model in the number of charged pocket residues that captures this dependence and reduces the mean absolute error from \SI{5.7}{\kilo\calorie\per\mole} to \SI{1.2}{\kilo\calorie\per\mole} against the reference in leave-one-out cross-validation.
For the within-target ranking, we find that Martini ranks congeneric ligands well when their affinity variation is driven by hydrophobicity, $\pi$-stacking, or ligand size, and poorly when it is driven by hydrogen bonding or ligand charge.
The structure--activity relationship of a series therefore provides a simple, interpretable predictor of whether Martini can accurately rank its ligands.
Both trends follow from the physics behind Martini's parametrization, which is calibrated against bulk partitioning data and therefore captures hydrophobic effects well but reproduces neither the directionality of hydrogen bonds nor the dielectric environment of a dense binding pocket.
Even with these limitations, Martini shares the statistical-mechanical basis of atomistic alchemical methods and yields free energies on a physically meaningful scale.
This places it in a different regime from MMGBSA, which is likewise used as a low-cost alternative to atomistic alchemical free-energy calculations.
MMGBSA retains the atomistic detail needed to rank congeneric ligands more accurately, but does not yield a proper free energy.
The two methods are therefore complementary: MMGBSA orders closely related ligands, whereas Martini compares chemically diverse molecules on a common scale.
Additionally, the same coarse resolution that blurs the chemical detail maps many atomistically distinct molecules onto one representation.
A single simulation therefore covers a whole neighborhood of chemical space.

Our assessment could be extended to more challenging systems such as GPCRs, macrocyclic ligands, and cofactor-containing targets.
Moreover, tuning the protocol for each target could modestly improve the accuracy of the free energies, although it would still be limited by the underlying Martini parametrization.
Our \emph{t4lys} results with the open-beta parameters indicate that the underbinding could be reduced by a different calibration of the non-bonded interactions.
However, a Martini variant fine-tuned for protein--ligand binding would reduce the transferability central to the model, whereas our pocket-based correction removes the offset without modifying the force field.
Beyond such model refinements, we see membrane proteins as a promising application, since Martini is well established in membrane environments\cite{Pedersen2025MartiniLipidome,Marrink2023TwoDecadesMartini} and this regime remains considerably more difficult for atomistic methods.

%% file: appendix.tex
\section*{\sffamily\bfseries\Large Supporting Information}
\label{sec:si}

This Supporting Information provides additional details on the methods and results described in the main text.
Its structure roughly mirrors that of the main text, i.e., methodological details, such as the simulation setup and ligand parametrization, are presented first (\cref{si:methods}), followed by supplementary results (\cref{si:results}).
\cref{si:ligandTopologies} illustrates the CG ligands.

\section{Methods}
\label{si:methods}

\subsection{Protein Targets}
\label{si:methods:proteinTargets}

\newcommand{\pcolwdt}[1]{\dimexpr #1\textwidth - 2\tabcolsep\relax}
\begin{table}[H]
    \renewcommand{\arraystretch}{1.1}
    \centering
    \caption{Overview of the 15 studied protein targets. Bold abbreviations are used throughout the manuscript. The number of ligands per target together with a short description of the protein is provided. Ligand details are given in \cref{si:ligandTopologies}.}
    \label{tab:si:proteinTargets}
    \begin{tabularx}{\linewidth}{
        >{\bfseries}l
        r
        >{\raggedright\arraybackslash}p{\pcolwdt{0.25}}
        >{\raggedright\arraybackslash}X
        }
        \toprule
        Abbrev.           & \# lig. & Protein Name                                & Description                                                                                                                                                                                                                                              \\
        \midrule
        \textbf{mup1}     & 6       & Major Urinary Protein I                     & Member of the MUP complex found in mice. Involved in pheromone binding and release. Naturally binds fragment sized ligands.\cite{Steinbrecher2015FragmentBFE}                                                                                            \\
        \addlinespace
        \textbf{mcl1}     & 12      & Myeloid Cell Leukemia 1                     & Member of the Bcl-2 protein family. Involved in cell death regulation and often overexpressed in human cancer. Interesting target for cancer research.\cite{Steinbrecher2015FragmentBFE}                                                                 \\
        \addlinespace
        \textbf{hsp90}    & 12      & Heat Shock Protein 90                       & Member of the HSP molecular chaperone and stress response family. One of the most abundant proteins in the cell. Relevant as anti-cancer drug target since it helps folding of proteins often overexpressed in cancer.\cite{Steinbrecher2015FragmentBFE} \\
        \addlinespace
        \textbf{t4lys}    & 12      & T4 Lysozyme L99A                            & Mutated form of the Lysozyme protein from the Enterobacteria Phage T4. The mutation creates an internal cavity able to bind small molecules. Widely used test system.\cite{Steinbrecher2015FragmentBFE}                                                  \\
        \addlinespace
        \textbf{thrombin} & 11      & Thrombin                                    & Enzyme that converts fibrinogen to fibrin, leading to blood clotting.\cite{UniProt2025}                                                                                                                                                                  \\
        \addlinespace
        \textbf{jak2}     & 12      & Janus Kinase 2                              & Member of the family of intracellular tyrosine kinases. Relevant for cell signaling. Potential therapeutic target in myeloproliferative disorders.\cite{Steinbrecher2015FragmentBFE}                                                                     \\
        \addlinespace
        \textbf{jnk1}     & 12      & Mitogen-activated protein kinase 8          & Member of the c-Jun N-terminal kinases. Involved in the regulation of gene expression in response to stress and inflammatory signals.\cite{UniProt2025}                                                                                                  \\
        \addlinespace
        \textbf{hif2a}    & 12      & Endothelial PAS domain-containing protein 1 & Transcription factor relevant for oxygen regulated genes.\cite{UniProt2025}                                                                                                                                                                              \\
        \addlinespace
        \textbf{cdk8}     & 12      & Cyclin Dependent Kinase 8                   & Activates and inhibits transcription by phosphorylation of other proteins.\cite{UniProt2025}                                                                                                                                                             \\
        \addlinespace
        \textbf{liga}     & 11      & Bacterial DNA Ligase                        & Enzyme responsible for fusing ends of DNA strands and relevant for bacterial growth. Interesting target for antibiotic research since it differs from human DNA ligases.\cite{Steinbrecher2015FragmentBFE}                                               \\
        \addlinespace
        \textbf{p38}      & 12      & p38$\alpha$ MAP Kinase                      & Member of the family of serine/threonine specific kinases. Relevant for mediating cellular stress response. Relevant therapeutic target in inflammatory diseases.\cite{Steinbrecher2015FragmentBFE}                                                      \\
        \addlinespace
        \textbf{eg5}      & 12      & Kinesin-like protein KIF11                  & Motor protein required for mitosis.\cite{UniProt2025}                                                                                                                                                                                                    \\
        \addlinespace
        \textbf{bace}     & 12      & Beta-secretase 1                            & Enzyme that cleaves the amyloid precursor protein. Relevance in Alzheimer's disease research.\cite{UniProt2025}                                                                                                                                          \\
        \addlinespace
        \textbf{cdk2}     & 12      & Cyclin-dependent kinase 2                   & Serine/threonine-protein kinase involved in cell cycle regulation.\cite{UniProt2025}                                                                                                                                                                     \\
        \addlinespace
        \textbf{cmet}     & 12      & Hepatocyte Growth Factor Receptor           & Transmits signals from the cell surface to the interior of the cell by binding the hepatocyte growth factor.\cite{UniProt2025}                                                                                                                           \\
        \bottomrule
    \end{tabularx}
\end{table}

\begin{table}[H]
    \renewcommand{\arraystretch}{1.1}
    \centering
    \caption{Composition of the 15 targets and their ligand series, ordered as in \cref{tab:si:proteinTargets}.
        $n$ is the number of ligands and the protein size the number of standard amino acids in the prepared structure, covering all chains of the simulated complex.
        The pocket columns give the number of pocket residues, the number of charged pocket residues, the fractions of charged, non-polar, aromatic, and polar residues, and the mean Kyte--Doolittle hydropathy of the pocket,\cite{Kyte1982Hydropathy} each averaged over the ligands of the target (\cref{si:methods:pocketAndLigandDescriptors}).
        The ligand columns give the ranges of heavy-atom counts and predicted logP values (\cref{sec:methods:ligands}) and the formal charges occurring in the series.}
    \label{tab:si:targetProperties}
    \input{si-target-data}
\end{table}

\subsection{Ligand Pocket Restraints}
\label{si:methods:restraints}
Similar to atomistic double-decoupling free-energy simulations, we used a pocket restraint to ensure a well-defined bound state.
While in atomistic simulations the restraint is typically switched off in the fully coupled state, or chosen such that it does negligible work, the faster dynamics of the CG simulations require the restraint to stay active to prevent the ligand from leaving the pocket even in the fully coupled state.
The faster dynamics also allow us to use a less tight restraint, which helps to avoid biasing the bound state.
Atomistic protocols instead commonly use Boresch-style restraints,\cite{Boresch2003Restraints} which fix the position and orientation of the ligand relative to the protein.
Since the faster CG dynamics sample the ligand orientation within the simulation, this restriction is not needed here.
Moreover, they require an accurate binding pose and a ligand-specific choice of reference beads, which is impractical to automate across a large and chemically diverse set of complexes.
We used a flat-bottom restraint,
\begin{equation}
    \label{eq:flatBottomRestraint}
    U_{\mathrm{restr}}(d)=\begin{cases}
        \frac{1}{2}k_{\mathrm{restr}}(d-r_{\mathrm{restr}})^2, & d \geq r_{\mathrm{restr}} \\
        0,                                                     & d < r_{\mathrm{restr}}
    \end{cases}
\end{equation}
where $d$ is the distance between the center of mass of the ligand and the pocket center, $r_{\mathrm{restr}}=\SI{0.4}{\nano\meter}$ the flat-bottom radius, and $k_{\mathrm{restr}}=\SI{2000}{\kilo\joule\per\mole\per\nano\meter\squared}$ the force constant.
The restraint was realized with the flat-bottom geometry of the GROMACS pull code and was applied in all $\lambda$ windows of the protein leg.
We implemented the pocket center as a non-interacting dummy particle, modeled as a virtual site whose position is constructed from four to eight (depending on the target) backbone beads of the protein.
The target position of the pocket center was specified by the ligand center of mass in the crystal structure of the protein-ligand complex from the dataset by \citeauthor{Baumann2026OpenFE}
The up to eight reference beads for the virtual site were chosen from all backbone beads within \SI{1.1}{\nano\meter} of the atomistic crystal ligand center of mass such that the sum of the pairwise distances between the reference beads is maximized, thereby ensuring that the reference beads are well spread out around the ligand.
We then optimized the weights of the reference beads with a constrained quadratic program that places the virtual site as close as possible to the reference ligand center of mass.
Concretely, we minimized the squared distance between the weighted average of the reference bead positions and the crystal ligand center of mass, $\lVert \sum_i p_i \mathbf{x}_i - \mathbf{r}_\mathrm{lig} \rVert^2$, with non-negative weights ($p_i \ge 0$) that sum to unity ($\sum_i p_i = 1$).
These constraints make the virtual site a convex combination of the reference beads, so that it always lies within their convex hull.
To prevent the weights from collapsing onto a single bead, we added an $\ell_2$ regularization term of strength $0.005$ that penalizes deviations of the weights from a uniform distribution (equivalently, their variance), favoring an even spread across the reference set.
The resulting problem was solved with sequential least-squares quadratic programming (SLSQP).
Since the virtual site is reconstructed from the backbone beads at every step, the pocket center follows the translation and rotation of the protein, and no additional restraints on the protein itself are required.
Defining the pocket center through a single dummy bead is more robust than defining it directly via distances to the individual reference beads, since the weighted average damps the fluctuations of the individual backbone beads and remains well defined even if part of the reference region is locally distorted.

The restraint also defines the sampled volume in the fully decoupled state of the alchemical transformation in the protein leg, which bounds the configurational volume of the near-decoupled windows and thereby preserves their phase-space overlap.\cite{Deng2009BindingFreeEnergies}
To make the free-energy result independent of the accessible restraint volume, we computed an analytical standard-state correction.
The free-energy correction is given by
\begin{equation}
    \label{eq:standardStateCorrection}
    \Delta G_{\mathrm{std}} = -k_\mathrm{B}T \ln \frac{V_{\mathrm{restr}}}{V^{\circ}}
    \qquad\text{with}\qquad
    V_{\mathrm{restr}} = \int_0^{\infty} 4\pi r^2 e^{-\beta U_{\mathrm{restr}}(r)} \mathrm{d}r
\end{equation}
where $V^{\circ} = 1/c^{\circ} = 1/(\SI{1}{\molar}) = \SI{1.6605}{\nano\meter\cubed}$ is the standard-state reference volume and $V_{\mathrm{restr}}$ the restraint volume.
Since the ligand is restrained to a single dummy particle, the accessible volume in the decoupled state is spherically symmetric and the integral can be evaluated in closed form, which would not be the case for a set of distance restraints to several individual reference beads.
For our setup with $r_{\mathrm{restr}}=\SI{0.4}{\nano\meter}$ and $k_{\mathrm{restr}}=\SI{2000}{\kilo\joule\per\mole\per\nano\meter\squared}$, we obtained a restraint volume of $V_{\mathrm{restr}} = \SI{0.37}{\nano\meter\cubed}$ and a standard-state correction of $\Delta G_{\mathrm{std}} = \SI{0.895}{\kilo\calorie\per\mole}$ at \SI{300}{\kelvin}.
The restraint volume is about \SI{38}{\percent} larger than the volume of the flat-bottom sphere, $\frac{4}{3}\pi r_{\mathrm{restr}}^3 = \SI{0.268}{\nano\meter\cubed}$, since the ligand can also sample the harmonic part of the restraint potential.
The sign here is defined such that the correction needs to be added to the free energy computed from the difference of the alchemical transformations, as written in \cref{eq:abfe} of the main text.

Since we also used $r_{\mathrm{restr}}$ as a definition of the bound state, we computed a bound state correction $\Delta G_{\mathrm{restr}}^{\mathrm{prot}}$ to account for the sampled configurations that lie outside the flat-bottom region, i.e., in the non-zero part of the restraint potential $U_{\mathrm{restr}}$.
This amounts to reweighting the sampled ensemble to the bound state, which gives $\Delta G_{\mathrm{restr}}^{\mathrm{prot}} = k_\mathrm{B}T \ln(\rho)$, where $\rho$ is the fraction of sampled configurations in the bound state, i.e., with $d < r_{\mathrm{restr}}$.
We determined $\rho$ from the fully coupled state only, since this is the state that defines the bound ensemble.
Since $\rho \leq 1$, the correction is negative, and with the sign convention of \cref{eq:abfe}, where it is subtracted, it slightly weakens the predicted binding: configurations in which the ligand pushes against the restraint wall lie outside the definition of the bound state and are removed from the bound ensemble.
Ideally, the bound state correction is small, since the restraint needs to be chosen such that it does not bias the bound state.
A large bound state correction therefore indicates that the flat-bottom radius is too small or that the restraint is not centered correctly.
For our simulations, we found that the bound state correction is indeed small for all targets, with a median of $\Delta G_{\mathrm{restr}}^{\mathrm{prot}} = \SI{-0.12}{\kilo\calorie\per\mole}$.

\subsection{Atomistic Reference Simulations}
\label{si:methods:atomisticReferenceSimulations}
We used atomistic simulations of the ligand in water to obtain reference distributions for the bonded degrees of freedom.
These distributions were used to derive bond, angle, and dihedral equilibrium values and force constants (\cref{si:methods:ligandParametrization}).
The atomistic simulations were performed with the OPLSAA force field\cite{Robertson2015OPLSAA} and GROMACS 2025.3.\cite{Abraham2025GROMACS}
Ligands were parametrized with the LigParGen web server.\cite{Dodda2017LigParGenServer}
Ligands were placed in a $(\SI{3.8}{\nano\meter})^3$ box and were solvated with TIP3P water using \texttt{gmx solvate}.\cite{Abraham2025GROMACS}
After a minimization stage, the system was equilibrated for \SI{250}{\pico\second}.
Next, the system was simulated for \SI{10}{\nano\second} with a time step of \SI{2}{\femto\second} in an NPT ensemble at \SI{298.15}{\kelvin} and \SI{1}{\bar}.

\subsection{Ligand Parametrization}
\label{si:methods:ligandParametrization}
The atom-to-bead mapping was defined manually with a custom tool inspired by CGBuilder,\cite{CGBuilder} in which atoms are assigned to beads by clicking them in the molecular structure.
Hydrogens follow their heavy atom automatically and atoms can be shared between beads.
For each bead the tool derives a signature of its chemical environment from the elements it contains, their number of bonds and hydrogens, their aromaticity, and their charge.
Each signature is associated with a bead type, which we took mostly from the parametrizations collected in the Martini bible.\cite{Martini3Bible}
Fragments with the same chemical environment therefore receive the same bead type across all ligands.

The tool also proposes the bonded topology.
Two beads are bonded if they are connected by an atomic bond, if they share an atom, or if they are bridged by a common atomic neighbor.
Of these, the bonds that arise from a shared atom or from an aromatic atomic bond are turned into constraints.
Beads with four or more constraints are turned into virtual sites, selected such that no two virtual sites are adjacent.
Their neighbors then form a rigid frame of constraints with the virtual site inside, which replaces a large constraint network by a small one, as established for fused rings in Martini small molecules.\cite{Alessandri2022MartiniSmallMolecules}
Angles and dihedrals are assigned to the remaining degrees of freedom, keeping only one angle per pair of outer beads and one proper dihedral per rotatable bond between two rigid units.
This prevents the same degree of freedom from being restrained by several potentials, which would over-stiffen the molecule.

The bonded parameters were then fitted to the mapped atomistic reference simulations with Fast-Forward,\cite{Brasnett2026FastForward, FastForward} for which we made three modifications.
First, we raised the force constant above which a bond is converted into a constraint, so that no bond is converted automatically.
Constraints were instead assigned in the mapping step, which keeps the constraint networks small enough to remain stable at a time step of \SI{20}{\femto\second}.
Second, we replaced the proper and improper dihedral terms by a combined bending-torsion potential.\cite{Bulacu2005BendingTorsionRigidity, Bulacu2013ImprovedAnglePotentials}
Its coefficients were obtained by Boltzmann inversion of the mapped dihedral distribution followed by a weighted least-squares fit, which penalizes the highest order terms to avoid overfitting the poorly sampled tails.
Since GROMACS multiplies the torsion by $\sin^3$ of the two adjacent bending angles at run time, we divided the coefficients by this prefactor evaluated at the mean bending angles, so that the fitted potential of mean force is recovered at the typical geometry.
This potential reproduced the atomistic dihedral distributions slightly less accurately than the standard proper and improper terms, but proved considerably more robust in our simulations.
Third, we removed the empirical cap of \SI{150}{\kilo\joule\per\mole\per\radian\squared} on the angle force constants, as stronger angle forces did not introduce instabilities in our systems.

After the fit, we capped the bond force constants at \SI{25000}{\kilo\joule\per\mole\per\nano\meter\squared} and wrote all angles with the cosine-based angle potential, since the harmonic and restricted bending terms assigned by Fast-Forward led to unstable simulations.
Angles with a fitted equilibrium value above \SI{175}{\degree} were instead replaced by a linear angle potential with a force constant of \SI{2000}{\kilo\joule\per\mole\per\nano\meter\squared},\cite{vanDerSpoel2020LinearMoieties} since an angle potential is ill defined at a collinear geometry.

\subsection{Ligand Validation Metrics}
\label{si:methods:validation}
The solvent accessible surface area was computed with \texttt{gmx sasa}\cite{Abraham2025GROMACS} for the isolated ligand, once for the CG and once for the atomistic model.
Both calculations used a probe radius of \SI{0.191}{\nano\meter} and 4800 test points per particle.
The atomistic radii were taken from \citeauthor{Rowland1996AtomRadii},\cite{Rowland1996AtomRadii} whereas the CG radii were assigned by bead size as \SI{0.264}{\nano\meter}, \SI{0.230}{\nano\meter}, and \SI{0.191}{\nano\meter} for regular, small, and tiny beads.\cite{Alessandri2022MartiniSmallMolecules}
Both structures were energy minimized before the SASA calculation.

Since experimental octanol--water partitioning data are not available for the studied ligands, we used predicted logP values as a reference.
Although these predictions are only approximate, they still provide a useful reference for the validation of the CG ligand models.
We averaged the four predictors into a consensus value and took the standard error of this mean as an uncertainty, which measures the disagreement between the methods rather than their accuracy.
Since logP values are not meaningful for charged molecules, we compared the partitioning of the neutralized ligand models (\cref{sec:methods:ligands}).
The CG counterpart was obtained by decoupling the ligand separately in water and in wet octanol, using the settings of the water leg with 20 $\lambda$ windows of \SI{2.4}{\nano\second} each.
In both phases the ligand was placed in a cubic box with a distance of \SI{4.5}{\nano\meter} between the ligand and its periodic images.
No ions were added.
The octanol phase was modeled with the Martini 1-octanol model at a water mole fraction of \SI{0.25}{}.
Since the neutralized ligand models carry no partial charges, only the van der Waals (vdW) interactions have to be decoupled.
The transfer free energy is the difference of the two decoupling free energies, with the sign defined such that positive values indicate hydrophobic ligands.

The bonded distributions of the CG ligands were taken from the fully coupled window of the partitioning simulation in water.
The atomistic reference distributions were obtained by mapping the atomistic trajectory to the CG resolution.
Constraints and virtual sites were excluded, since their geometry is fixed by construction.
Each remaining bond, angle, and dihedral was then scored against its reference with the \texttt{ff\_assess} metric, which combines the Hellinger distance between the two distributions with the difference of their means.\cite{Brasnett2026FastForward}
The mean difference is normalized by twice the standard deviation of the reference and capped at unity, and the two terms enter with weights of \SI{0.7}{} and \SI{0.3}{}.
The resulting score lies between 0 and 1, with lower values indicating better agreement.
We report the mean score over all interactions of a ligand, separated by interaction type in \cref{fig:si:bondedScoreDistribution}.

\subsection{Simulation Details}
\label{si:methods:simulationDetails}
For the protein leg, the CG complex was placed in a cubic box with a minimum distance of \SI{2.1}{\nano\meter} between the protein and the box edge and solvated with regular Martini water using \texttt{insane}.\cite{Wassenaar2015Insane}
For the water leg, the ligand was placed in a cubic box with a distance of \SI{4.5}{\nano\meter} between the ligand and its periodic images.
In both legs we added \SI{0.15}{\molar} NaCl and neutralized the system with additional counter ions.

All simulations used the stochastic dynamics integrator with a time step of \SI{20}{\femto\second}, an inverse friction constant of \SI{2}{\pico\second}, and a reference temperature of \SI{300}{\kelvin}.
The pressure was kept at \SI{1}{\bar} with the isotropic c-rescale barostat,\cite{Bernetti2020CRescale} using a coupling time of \SI{4}{\pico\second} and a compressibility of \SI{3e-4}{\per\bar}.
Non-bonded interactions were treated with the Verlet scheme, a Verlet list radius of \SI{1.35}{\nano\meter} updated every 20 steps, and a cutoff of \SI{1.1}{\nano\meter} for both the potential-shifted Lennard-Jones and the reaction-field electrostatics.\cite{Kim2023NeighborListArtifacts}
Constraints were solved with LINCS using an expansion order of 8 and two iterations, as recommended for the large time step of the CG model.
Each $\lambda$ window was energy minimized with up to 30000 steepest descent steps and equilibrated for \SI{0.8}{\nano\second} with velocities generated at \SI{300}{\kelvin}.
During the equilibration of the protein leg, a harmonic restraint with a force constant of \SI{2000}{\kilo\joule\per\mole\per\nano\meter\squared} held the ligand at the pocket center, which was replaced by the flat-bottom restraint of \cref{eq:flatBottomRestraint} in the production run.

The production runs covered \SI{50}{\nano\second} per $\lambda$ window in the protein leg and \SI{2.4}{\nano\second} per window in the water leg.
The ligand was decoupled from its environment and from its own non-bonded interactions, with soft-core interactions of $\alpha = 0.5$, $p = 1$, and $\sigma = \SI{0.3}{\nano\meter}$ to avoid singularities in the end states.
Since both legs used the same setting, the intramolecular contribution cancels in the binding free energy.
The energies of all $\lambda$ states were written every \SI{2}{\pico\second} and served as input for the MBAR estimate.
The same protocol was applied to all 15 targets and 172 ligands, without tuning the simulation settings for individual systems.

\subsection{Comparison of runtime performance with OpenFE}
\label{si:methods:runtimeComparison}
We compared the runtime against OpenFE rather than against our main FEP+ reference, since \citeauthor{Ross2023SchroedingerFEP} do not report runtime information.\cite{Ross2023SchroedingerFEP}
All timings are reported in GPU hours, i.e., the wall time of a simulation multiplied by the number of GPUs it occupied.
For Martini, each replica-exchange simulation ran all $\lambda$ windows of one leg simultaneously on two NVIDIA A30 GPUs, and we took its wall time from the GROMACS log of the first window.
The cost per ligand sums the three repeats of the protein leg and the two repeats of the water leg.
For OpenFE, we took the total run time of each simulation from the dataset of \citeauthor{Baumann2026OpenFE}, which ran all $\lambda$ windows of one leg on a single NVIDIA L40S GPU.\cite{Baumann2026OpenFE}
Its peak single-precision throughput of \SI{91.6}{} TFLOPS exceeds the \SI{10.3}{} TFLOPS of the A30 used here, while the memory bandwidths are comparable (\SI{864}{} against \SI{933}{\giga\byte\per\second}).\cite{NvidiaL40S,NvidiaA30}
Since their networks contain more edges than ligands, the cost per ligand divides the total cost of a network, including all edges, both legs, and three repeats each, by its number of ligands.
The cost per ligand thus averages over all ligands of a target rather than only those selected for this study, since the edge network makes extracting the cost of individual ligands difficult.
One repeat of the protein leg takes on average \SI{5.6}{} GPU hours per charge-conserving edge, but \SI{63}{} GPU hours for the six charge-changing edges.

\begin{table}[H]
    \renewcommand{\arraystretch}{1.1}
    \centering
    \caption{Runtime comparison between the Martini protocol of this study and the atomistic OpenFE reference calculations, in GPU hours. The protein and water columns give the cost of a single repeat, per edge for OpenFE and per ligand for Martini, as the mean over the edges or ligands of a target. The per-ligand columns give the full cost of one binding free energy, including all repeats.}
    \label{tab:si:runtimeComparison}
    \begin{tabular}{>{\bfseries}l rrr rrr}
        \toprule
                 & \multicolumn{3}{c}{OpenFE} & \multicolumn{3}{c}{Martini}                                         \\
        \cmidrule(lr){2-4} \cmidrule(lr){5-7}
        Target   & protein                    & water                       & per lig. & protein & water & per lig. \\
        \midrule
        mup1     & 1.90                       & 0.57                        & 8.65     & 0.72    & 0.01  & 2.19     \\
        mcl1     & 2.71                       & 0.82                        & 14.91    & 0.81    & 0.02  & 2.48     \\
        hsp90    & 2.58                       & 0.58                        & 10.94    & 0.87    & 0.02  & 2.73     \\
        t4lys    & 3.16                       & 0.76                        & 13.74    & 0.84    & 0.01  & 2.63     \\
        thrombin & 3.76                       & 0.81                        & 17.45    & 1.03    & 0.02  & 3.23     \\
        jak2     & 4.58                       & 0.83                        & 23.47    & 1.23    & 0.02  & 3.94     \\
        jnk1     & 5.11                       & 0.58                        & 21.93    & 1.38    & 0.02  & 4.19     \\
        hif2a    & 2.26                       & 0.66                        & 12.63    & 0.79    & 0.02  & 2.53     \\
        cdk8     & 10.05                      & 0.88                        & 52.27    & 2.27    & 0.02  & 7.04     \\
        liga     & 7.99                       & 0.68                        & 30.71    & 1.35    & 0.02  & 4.11     \\
        p38      & 7.10                       & 0.64                        & 34.84    & 1.31    & 0.02  & 4.29     \\
        eg5      & 15.51                      & 1.32                        & 77.54    & 1.45    & 0.02  & 4.77     \\
        bace     & 6.04                       & 0.60                        & 27.08    & 1.55    & 0.02  & 5.10     \\
        cdk2     & 10.77                      & 0.75                        & 51.85    & 1.40    & 0.02  & 4.25     \\
        cmet     & 6.01                       & 1.04                        & 34.37    & 1.25    & 0.02  & 3.80     \\
        \midrule
        Mean     & 5.97                       & 0.77                        & 28.82    & 1.22    & 0.02  & 3.82     \\
        \bottomrule
    \end{tabular}
\end{table}

\subsection{Optimization of the \texorpdfstring{$\lambda$}{lambda} Schedule}
\label{si:methods:lambdaSchedule}
Both the efficiency of the Hamiltonian replica exchange and the precision of the MBAR estimate are limited by the pair of adjacent $\lambda$ windows with the smallest phase-space overlap.
An efficient schedule therefore spaces the windows for uniform overlap along the alchemical path, rather than spacing $\lambda$ uniformly.
We used the same schedule for all ligands, optimized based on a proposal schedule that was evaluated on a subset of ligands.
We started from a manually constructed schedule with a linearly spaced electrostatic leg and a non-linearly spaced vdW leg.
A purely linear proposal would serve equally well, since the refinement only requires it to resolve the path sufficiently.
During the proposal simulations we computed the energy of every sampled configuration in every $\lambda$ state, so the exchange acceptance can be predicted for any pair of states.
We used this acceptance as the measure of overlap, since it directly limits the replica exchange and requires no additional simulations.
For each pair we took the lowest acceptance across the evaluated ligands.
This makes the schedule robust for different ligands.
To translate overlap into window positions, we assumed Gaussian distributed exchange energies.
The acceptance of a pair then depends only on the distance between the two states, analogous to a thermodynamic length.
Summing these distances gives a cumulative length along the leg, and equally spaced windows in this length yield a uniform acceptance.
The refinement thus adds windows where the overlap of the proposal is poor and removes them where it is higher than necessary.
The vdW and electrostatic legs and the protein and water legs were each optimized separately.
Targeting an acceptance of \SI{0.35}{}, the electrostatic decoupling requires only 4 states in the protein leg and 3 states in the water leg.
The vdW decoupling uses 20 and 17 windows, with the spacing tightening towards the decoupled state where the overlap is hardest to maintain.

\subsection{Pocket and Ligand Descriptors}
\label{si:methods:pocketAndLigandDescriptors}
The pocket descriptors were computed on the atomistic structures prepared by \citeauthor{Baumann2026OpenFE} together with the crystal pose of the respective ligand.\cite{Baumann2026OpenFE}
We use the ligand structures to define the binding pocket, with all residues with a heavy atom within \SI{0.6}{\nano\meter} of any ligand heavy atom as pocket residues.
Since all ligands were placed in the same protein structure, the variation between the ligands reflects the ligand footprint only and not the protein conformation.
We classified the pocket residues as aromatic (Phe, Tyr, Trp, His), polar (Ser, Thr, Asn, Gln, Cys, Tyr, His, Asp, Glu, Lys, Arg), and non-polar (Ala, Val, Leu, Ile, Met, Pro, Gly, Phe, Trp).
The charge of a residue follows from the protonation states of the prepared structures, which we obtained by summing the Amber14 partial charges over the residue.
Residues with a non-zero net charge were assigned to the charged class only and are not counted as polar (Asp, Glu, Lys, Arg, His) or aromatic (His).

The ligand series descriptors combine intrinsic ligand properties with the interactions formed in the crystal pose.
The logP is the mean of the four predictions also used for the ligand validation (\cref{si:results:ligandValidation}), the size is the number of heavy atoms, and the charge is the formal charge of the ligand.
The ligand interactions were determined with PLIP\cite{Salentin2015PLIP} on a complex rebuilt from the prepared apo structure and the crystal ligand pose, covering hydrophobic contacts, hydrogen bonds, salt bridges, water bridges, $\pi$-stacking, and $\pi$-cation interactions.
Hydrogen bonds were additionally split into those donated by the ligand and those donated by the protein.
Since these counts come from a single structure, they describe the binding mode of the crystal pose and not an average over the simulation.
They nevertheless provide a useful descriptor of the interactions formed by the ligand series.

\section{Results}
\label{si:results}

\subsection{Ligand Validation}
\label{si:results:ligandValidation}
\begin{figure}[H]
    \centering
    \includegraphics[width=\columnwidth]{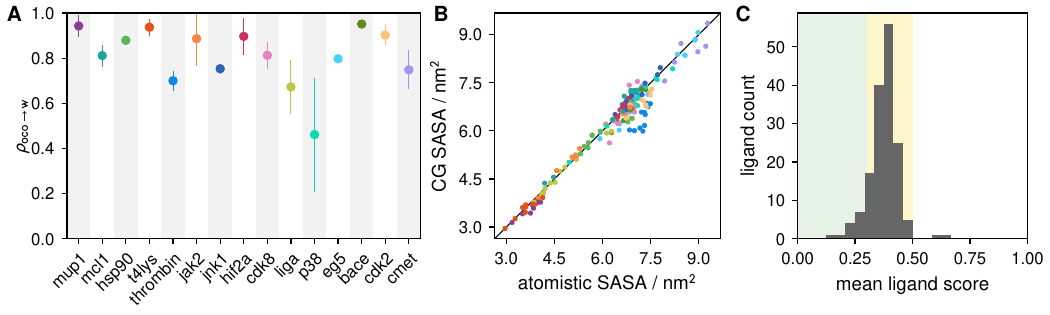}
    \caption{Overview of the ligand validation results for the three tested quantities.
        \textbf{(A)} Per target Spearman correlation for the octanol--water partitioning between the simulation results for the neutralized CG ligand models and the average over four cheminformatics logP predictions\cite{Tetko2002ALOGPS, Cheng2007XLOGP3, Moriguchi1992MLOGP, Lipinski2001MLOGP, SilicosIT} (\cref{fig:si:ocoWPartitioning}).
        Error bars indicate $\pm1$ standard error.
        The CG results correlate well with the reference for all targets, with a median Spearman correlation of \SI{0.81}{}.
        \emph{p38} with the worst correlation is also the target where the predictions disagree with each other the most.
        \textbf{(B)} Parity plot of the CG SASA values compared to the atomistic reference values.
        Colors indicate the target the ligand corresponds to (cf. panel A).
        The CG SASA values are in good agreement with the atomistic reference, with a slight tendency to underestimate the SASA.
        The mean absolute error is \SI{0.21}{\nano\meter\squared} and the median relative error is \SI{0.024}{}.
        The maximum relative error is \SI{0.174}{} with six ligands having a relative error above \SI{0.1}{}.
        Ligands from the \emph{thrombin} target show the largest deviations.
        \textbf{(C)} Histogram of the mean bonded distribution scores per ligand computed with \texttt{ff\_assess}.\cite{Brasnett2026FastForward}
        The green region (score $< 0.3$) indicates good agreement with the atomistic reference (15 ligands), while the yellow region (score $0.3$--$0.5$) indicates acceptable agreement (141 ligands).
        Only two ligands have a mean bonded distribution score above $0.5$.
        The remaining 14 ligands have no flexible bonded terms and therefore no score.}
    \label{fig:si:ligandValidation}
\end{figure}

\begin{figure}[H]
    \centering
    \includegraphics[width=\columnwidth]{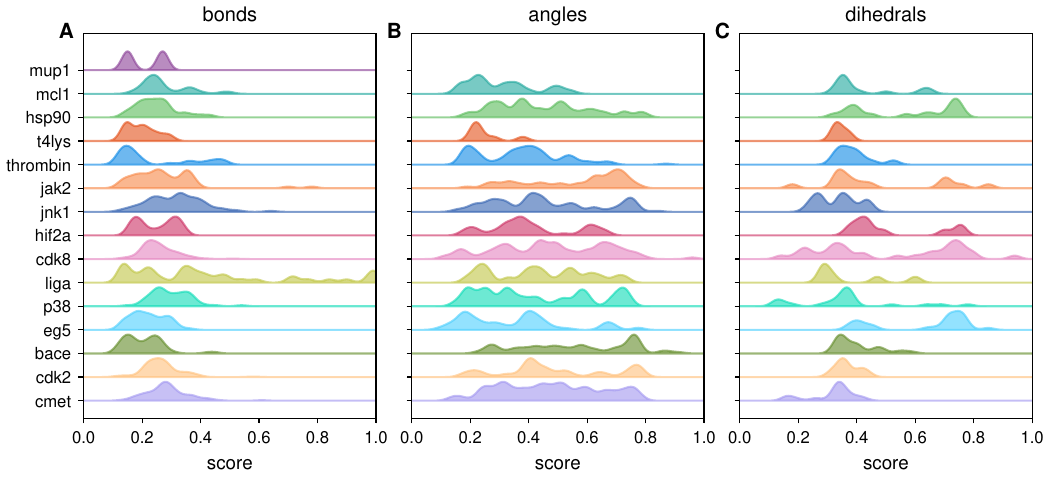}
    \caption{Per target distributions of the bonded distribution scores computed with \texttt{ff\_assess}\cite{Brasnett2026FastForward} for \textbf{(A)} bonds, \textbf{(B)} angles, and \textbf{(C)} dihedrals.
        Each curve is a kernel density estimate over all interactions of the respective type of all ligands of a target, scaled to equal height.
        Targets are colored as in \cref{fig:si:ligandValidation}A and ordered by ligand ranking correlation.
        \emph{mup1} ligands contain neither angle nor dihedral potentials.
        Bond scores are consistently low (median \SI{0.25}{}), while angles (median \SI{0.43}{}) and dihedrals (median \SI{0.37}{}) are broader and show tails towards larger scores.}
    \label{fig:si:bondedScoreDistribution}
\end{figure}

\begin{figure}[H]
    \centering
    \includegraphics[width=\columnwidth]{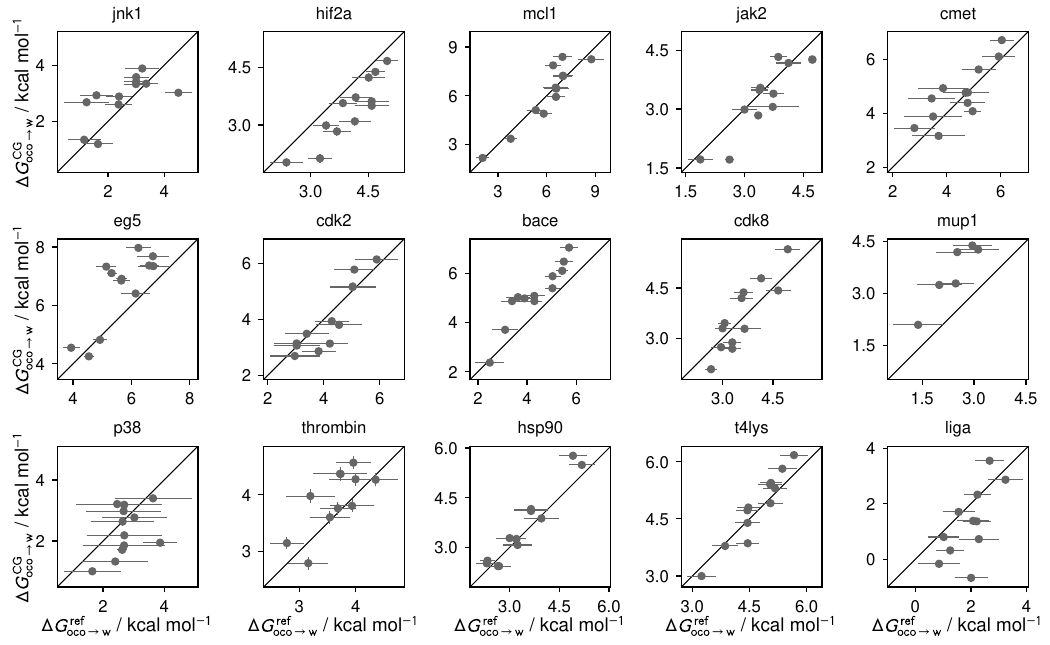}
    \caption{Octanol--water partitioning free-energy results for the neutralized CG ligand models compared with the averaged atomistic predictions.
        logP predictions are converted to free energies using $\Delta G_{\mathrm{ow}} = RT \ln(10) \, \mathrm{logP}$.
        The reference is the average over the logP predictions of ALOGPS,\cite{Tetko2002ALOGPS} XLOGP3,\cite{Cheng2007XLOGP3} MLOGP,\cite{Moriguchi1992MLOGP, Lipinski2001MLOGP} and SILICOS-IT\cite{SilicosIT} (\cref{sec:methods:ligands}).
        Error bars indicate $\pm1$ standard error of the mean of the four prediction methods.}
    \label{fig:si:ocoWPartitioning}
\end{figure}

Ligands were validated using three different metrics:
the octanol--water partitioning, the solvent accessible surface area (SASA), and the bonded distributions.
\cref{fig:si:ligandValidation} provides an overview of the ligand validation results for all three tested quantities.
\cref{fig:si:ocoWPartitioning,fig:si:bondedScoreDistribution} show a detailed per target analysis for the partitioning results and the bonded distribution scores, respectively.

The systematic underestimation of the SASA by the CG models likely arises from focusing on tiny and small beads to preserve chemical details, which reduces the effective SASA per atom compared.
An extrem example for this effect can be seen when mapping a straight dodecane chain with 6 tiny, 4 small, or 3 regular sized beads, yielding SASA values of \SI{5.00}{\nano\meter\squared}, \SI{5.41}{\nano\meter\squared}, and \SI{5.64}{\nano\meter\squared}, respectively, These values lie \SI{12}{\percent}, \SI{5}{\percent}, and \SI{1}{\percent} below the atomistic value of \SI{5.70}{\nano\meter\squared}, respectively.
Preserving chemical detail with smaller beads therefore comes at the cost of a lower surface area per atom.

\begin{figure}[H]
    \centering
    \includegraphics[width=\columnwidth]{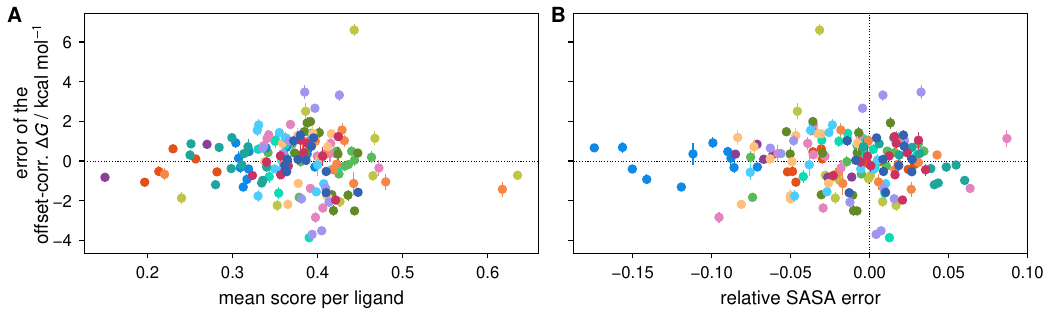}
    \caption{Error of the offset-corrected CG binding free energies against the bonded distribution scores and the relative SASA error of the ligands.
        The error is given as the deviation from the FEP+ reference after removing the per target underbinding $\Delta$.
        \textbf{(A)} Error against the mean bonded distribution score of the ligand (cf. \cref{fig:si:bondedScoreDistribution}).
        \textbf{(B)} Error against the relative SASA error (SASA$_\mathrm{CG}$/SASA$_\mathrm{AA}$ - 1) of the ligand (cf. \cref{fig:si:ligandValidation}B).
        Colors indicate the target and error bars indicate one standard error, combining the CG and reference free-energy uncertainties.
        Neither metric correlates with the error, indicating that the remaining inaccuracies of the ligand models do not propagate into the binding free energies.}
    \label{fig:si:residualVsLigandMetrics}
\end{figure}

To test whether lower accuracy in the ligand parametrization results in larger errors in the free energies, we compared the error of the target-offset-corrected binding free energy with the mean bonded distribution scores and the relative SASA error (\cref{fig:si:residualVsLigandMetrics}).
Both comparisons show no correlation, indicating that the small errors in the ligand parametrization do not significantly affect the binding free-energy results.

\subsection{Comparison with \texorpdfstring{\citeauthor{Souza2020ProteinLigandBinding}}{Souza et al.}}
\label{si:results:souzaComparison}
\citeauthor{Souza2020ProteinLigandBinding} computed absolute binding free energies for seven small ligands of T4 lysozyme from potentials of mean force computed from unbiased simulations.\cite{Souza2020ProteinLigandBinding}
Their results provide a useful reference for our protocol.
Since they used the open-beta version of Martini~3, which differs substantially from the official release, we recomputed their seven systems with our protocol and the open-beta parameters for both the ligands and the protein.
Benzene, for example, is built from three TC4 beads in the open beta and from three TC5 beads in the official Martini~3 release, with the definitions of the bead types also differing.
We also replaced the OLIVES G\={o}-model by an elastic network with a force constant of \SI{500}{\kilo\joule\per\mole\per\nano\meter\squared} and an upper cutoff of \SI{0.9}{\nano\meter} to stay close to the reference protein model.
The pocket restraint and the free-energy estimation are identical to the rest of our double-decoupling protocol.
We ran two independent repeats per leg.
We additionally ran the same systems with the official Martini~3 release and an elastic network model as the structure bias model.
Four of the ligands are also part of the main dataset, where they were simulated with the official release and the OLIVES G\={o}-model, which allows us to separate the effect of the non-bonded parameters from that of the structural bias.

\begin{figure}[H]
    \centering
    \includegraphics[width=\columnwidth]{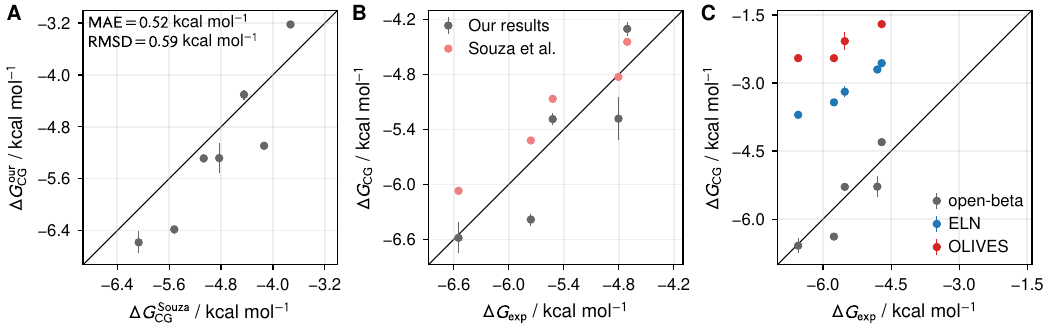}
    \caption{Comparison of absolute binding free energies for the seven T4 lysozyme L99A ligands of \citeauthor{Souza2020ProteinLigandBinding}
    \textbf{(A)} Our results with the open-beta Martini~3 parameters $\Delta G_{\mathrm{CG}}^{\mathrm{our}}$ against the reference values $\Delta G_{\mathrm{CG}}^{\mathrm{Souza}}$,\cite{Souza2020ProteinLigandBinding} together with the MAE and the root-mean-square deviation between the two.
    \textbf{(B)} Both sets of free-energy values against the experimental binding free energies, which are available for five of the seven ligands.
    The other two ligands do not bind experimentally and no experimental references are therefore available.
    \textbf{(C)} Our results against the experimental values, computed with the open-beta (gray) and with the official Martini~3 release parameters (blue), both using an elastic network (ELN), and with the official parameters and the OLIVES G\={o}-model (red).
    The latter are taken from the main dataset and only include four of the ligands evaluated by \citeauthor{Souza2020ProteinLigandBinding}
    Lines mark the diagonal and error bars indicate one standard error over the repeats.}
    \label{fig:si:souzaComparison}
\end{figure}

Our open-beta results reproduce the reference values with a mean absolute error (MAE) of \SI{0.52}{\kilo\calorie\per\mole} and a root-mean-square deviation of \SI{0.59}{\kilo\calorie\per\mole} (\cref{fig:si:souzaComparison}A).
Both sets of values agree comparably well with experiment, with a MAE of \SI{0.29}{\kilo\calorie\per\mole} for the reference and \SI{0.36}{\kilo\calorie\per\mole} for our results (\cref{fig:si:souzaComparison}B).
Our efficient double-decoupling protocol therefore reproduces the binding free energies of the underlying model.
With the official Martini~3 release, the ligands bind more weakly, while their ranking remains unchanged.
For the five ligands with an experimental reference, the open-beta results show no systematic offset, with a mean signed error of \SI{-0.10}{\kilo\calorie\per\mole}, whereas the release parameters underbind by \SI{2.35}{\kilo\calorie\per\mole} (\cref{fig:si:souzaComparison}C).
Since the two sets differ only in the non-bonded parameters and share the protocol, the protein model, and the ligand mapping, this offset is a property of the force field rather than of our protocol.
The structural bias model shifts it further: with the OLIVES G\={o}-model instead of the elastic network, the underbinding of these four ligands increases from \SI{2.41}{\kilo\calorie\per\mole} to \SI{3.46}{\kilo\calorie\per\mole}.
For \emph{t4lys}, the elastic network therefore appears better suited than the OLIVES model, probably because the latter is too flexible for the dry cavity inside the protein.

\subsection{Unbiased Simulations}
\label{si:results:unbiasedSimulationResults}
To validate our double-decoupling free-energy results, we performed unbiased simulations for five ligands across four targets.
The unbiased simulations were performed with the same setup as the double-decoupling simulations, but without any alchemical transformations.
Instead of the pocket restraint we used a flat-bottom restraint with a radius of \SI{3.5}{\nano\meter}.
This restraint does not restrain the ligand in the pocket, but restricts the volume accessible to the unbound ligand.
By reducing this accessible volume compared to the full simulation box, the probability of binding events is increased and the sampling efficiency is improved.
The volume dependence is resolved by a standard-state correction.

For each system we ran eight independent replicas of \SI{0.5}{} (for the \emph{t4lys} target) to \SI{1.5}{\micro\second} length.
Half of the replicas started from the bound pose used in the double-decoupling simulations and half from unbound structures in which the ligand is \SI{2.5}{} to \SI{3.2}{\nano\meter} away from the pocket.
Both sets therefore approach the equilibrium population from opposite sides, and the difference between their free energies serves as a convergence check.
In total, the five reported systems cover \SI{44}{\micro\second} of sampling over 40 replicas, in which we observed 1258 binding events.

We computed the binding free energy from the populations of the bound and the unbound state,
\begin{equation}
    \label{eq:unbiasedBindingFreeEnergy}
    \Delta G = -k_\mathrm{B}T \ln \frac{N_{\mathrm{bound}}}{N_{\mathrm{bulk}}} \frac{V_{\mathrm{bulk}}}{V^{\circ}}
\end{equation}
where $N_{\mathrm{bound}}$ is the number of frames with the ligand center of mass within \SI{0.4}{\nano\meter} of the pocket center, i.e., the same bound state definition as in the double-decoupling simulations.
$N_{\mathrm{bulk}}$ counts the frames in which all ligand beads are further than the interaction cutoff of \SI{1.1}{\nano\meter} away from any protein bead, so that the ligand is fully decoupled from the protein.
Frames in which the ligand is attached to the protein surface outside the pocket contribute to neither population and are therefore excluded, although they can make up a large fraction of the trajectories since the ligands tend to stick to the protein surface.
The free energy of \cref{eq:unbiasedBindingFreeEnergy} thus describes the transfer of the ligand from the bulk into the pocket, consistent with the bound state of the double-decoupling simulations.
$V_{\mathrm{bulk}}$ is the volume accessible to the ligand in this unbound region.
We determined it by inserting 4000 random test points into the confining sphere and recording their distance to the nearest protein bead.
The fraction of points beyond the bulk distance, averaged over the trajectory and multiplied by the volume of the confining sphere, gives $V_{\mathrm{bulk}}$ and thereby accounts for the volume occupied by the protein.
$V^{\circ}$ is the standard-state volume of \cref{eq:standardStateCorrection}.
For the uncertainty we resampled the replicas with replacement 200 times and recomputed the free energy for each bootstrap sample.

\begin{figure}[H]
    \centering
    \includegraphics[width=\columnwidth]{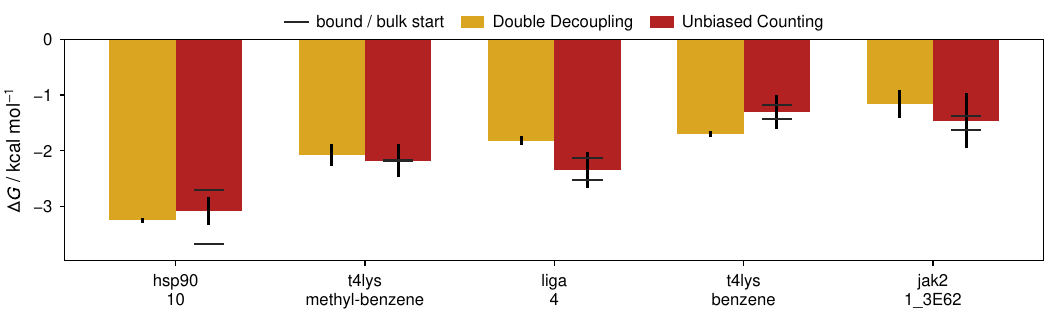}
    \caption{Binding free energies from unbiased simulations compared with the double-decoupling results for five ligands across four targets.
        Error bars give the standard error over the repeats for the double-decoupling results and the bootstrap uncertainty over the replicas for the unbiased results.
        Horizontal lines mark the free energies obtained separately from the replicas started in the bound pose and from bulk water, which bracket the pooled value and indicate the remaining start dependence.
        A larger spread between the two indicates limited convergence of the unbiased simulation results.
        Both protocols agree with a mean signed error (unbiased minus double decoupling) of \SI{-0.07}{\kilo\calorie\per\mole} and a MAE of \SI{0.30}{\kilo\calorie\per\mole}.}
    \label{fig:si:unbiasedVsTI}
\end{figure}

\subsection{Simulation Convergence}
\label{si:results:convergence}

\begin{figure}[H]
    \centering
    \includegraphics[width=\columnwidth]{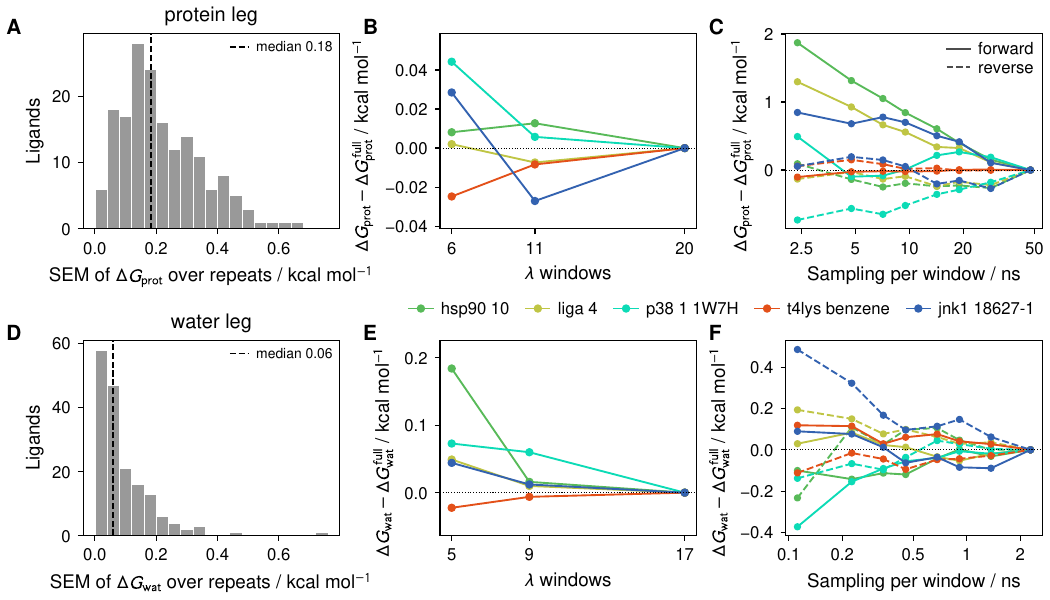}
    \caption{Convergence of the double-decoupling free energies for the protein leg \textbf{(A--C)} and the water leg \textbf{(D--F)}.
        \textbf{(A, D)} Distribution of the standard error of the decoupling free energy over the independent repeats for all 172 ligands (three repeats for the protein and two for the water leg).
        The dashed line marks the median of \SI{0.18}{\kilo\calorie\per\mole} and \SI{0.06}{\kilo\calorie\per\mole}, respectively.
        \textbf{(B, E)} Decoupling free energy of five ligands from five different targets when the $\lambda$ schedule is coarsened by removing intermediate windows, given relative to the result of the full schedule $\Delta G^{\mathrm{full}}$.
        \textbf{(C, F)} Same difference when only a fraction of the sampling of each window is used, taken either from the beginning (forward, solid) or from the end (reverse, dashed) of the trajectory.
        Halving the number of windows changes the free energies by at most \SI{0.06}{\kilo\calorie\per\mole} and using only \SI{60}{\percent} of the sampling by at most \SI{0.27}{\kilo\calorie\per\mole}.}
    \label{fig:si:convergence}
\end{figure}

\subsection{Sensitivity to the Flat-Bottom Radius}
\label{si:results:flatBottomRadius}
\begin{figure}[H]
    \centering
    \begin{minipage}[t]{0.5\columnwidth}
        {\sffamily\bfseries (A)}\par\vspace{2pt}
        \centering
        \includegraphics[width=\columnwidth]{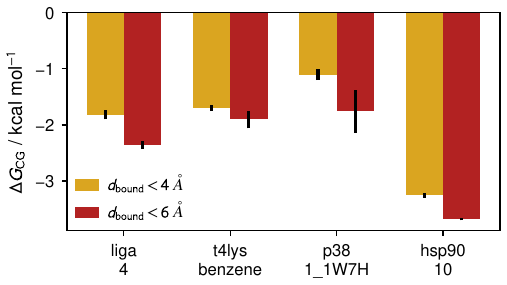}
    \end{minipage}
    \begin{minipage}[t]{0.49\columnwidth}
        {\sffamily\bfseries (B)}\par\vspace{2pt}
        \centering
        \includegraphics[width=0.9\columnwidth]{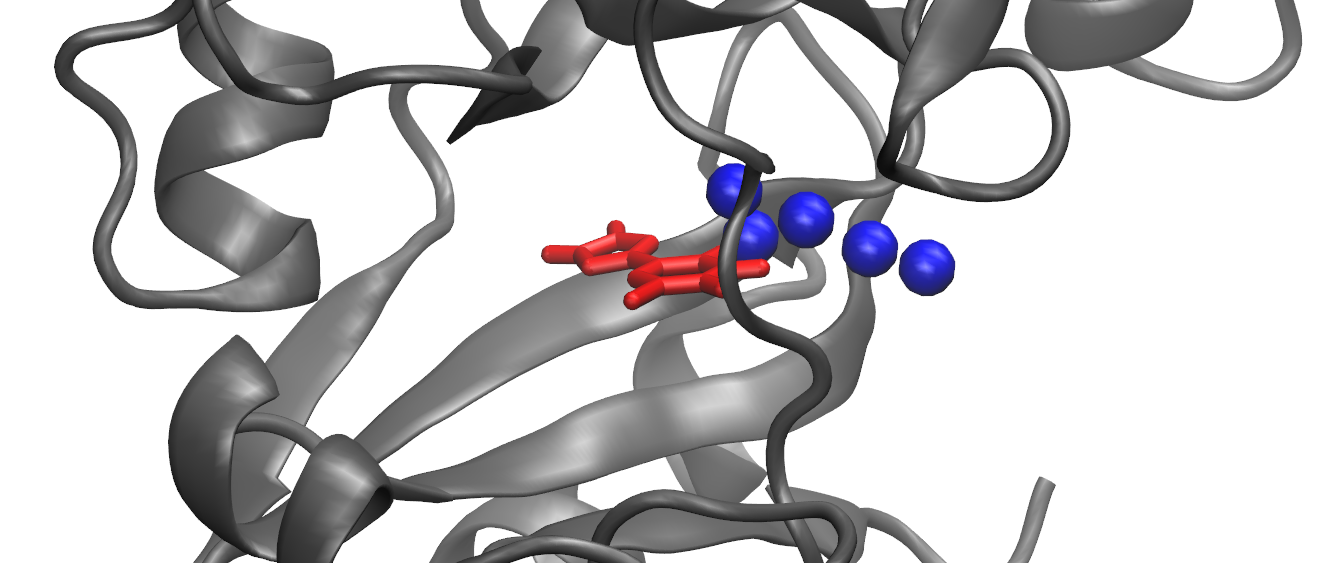}
    \end{minipage}
    \caption{Influence of the flat-bottom radius on the binding free energy.
        \textbf{(A)} Binding free energies of four ligands computed with a flat-bottom radius of \SI{0.4}{\nano\meter} and \SI{0.6}{\nano\meter}, which also sets the bound state cutoff $d_{\mathrm{bound}}$.
        The larger radius predicts stronger binding by \SI{0.45}{\kilo\calorie\per\mole} on average.
        Error bars indicate one standard error.
        \textbf{(B)} Snapshot of the \emph{liga} ligand 4 simulated with $r_{\mathrm{restr}}=\SI{0.6}{\nano\meter}$, showing the protein (grey), the CG ligand (blue), and the crystallographic pose of the atomistic ligand (red).
        The ligand center of mass is \SI{0.55}{\nano\meter} away from the pocket center and the ligand no longer occupies the native site, yet it is still counted as bound.}
    \label{fig:si:dG4VsDG6}
\end{figure}

Since $r_{\mathrm{restr}}$ defines the bound state, it was chosen tight enough to keep the CG ligand in the binding mode of the reference structure but loose enough to avoid biasing the bound state (\cref{si:methods:restraints}).
To probe this choice, we recomputed four ligands from four targets with an enlarged radius of \SI{0.6}{\nano\meter}.
The predicted binding becomes stronger by \SI{0.45}{\kilo\calorie\per\mole} on average (\cref{fig:si:dG4VsDG6}A).
The standard-state correction already accounts for the enlarged volume at the decoupled end, so a pocket that confines the ligand by itself would leave the binding free energy unchanged.
Uniform sampling of the enlarged volume would instead lower it by \SI{0.66}{\kilo\calorie\per\mole}.
The observed shift lies between these two limits, indicating that the restraint actively restricts the ligand rather than only defining the bound state.
However, this restriction is intended, since with the wider radius the ligand also samples configurations far from the native bound state (\cref{fig:si:dG4VsDG6}B).
Such configurations still count towards the bound state and could reduce the specificity of the binding free energy.
The weak confinement is consistent with the coarse and isotropic Martini interactions, which define the bound state less sharply than an atomistic model.
We therefore kept the tighter radius of \SI{0.4}{\nano\meter}, which ensures that the ligand samples the native bound state of the atomistic reference.

\subsection{Charged Ligands and the Electrostatics Treatment}
\label{si:results:offsetChargeTest}
We used the Reaction-Field method to treat electrostatics,\cite{Tironi1995GeneralizedReactionField} as typically done in Martini simulations.\cite{Souza2021Martini3}
The methods truncates electrostatic interactions at a cutoff of \SI{1.1}{\nano\meter} and replaces the longer-ranged part by a dielectric continuum.
Although less accurate then the Particle Mesh Ewald method,\cite{Darden1993PME} it simplifies the decoupling of charged ligands, which would otherwise require a counterion to maintain charge neutrality and avoid artifacts from the periodic boundary conditions.

\begin{figure}[H]
    \centering
    \includegraphics[width=0.6\columnwidth]{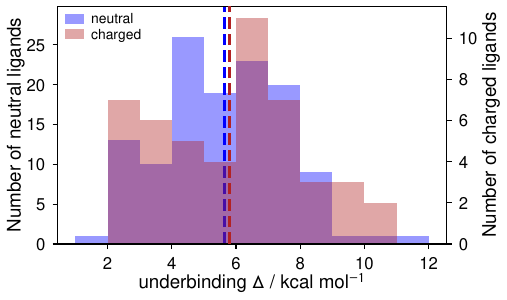}
    \caption{Underbinding $\Delta$ for the 124 neutral (left axis) and 48 charged (right axis) ligands.
        The two count axes are scaled such that equal fractions of each set have equal height, so the distributions can be compared directly.
        Dashed lines mark the respective means, which differ by \SI{0.18(38)}{\kilo\calorie\per\mole}.}
    \label{fig:si:chargeErrorHistogram}
\end{figure}

Of the 172 ligands, 48 carry a net charge of $\pm 1$, and their absolute deviation from the reference values differs from that of the 124 neutral ligands by only \SI{0.18(38)}{\kilo\calorie\per\mole} (\cref{fig:si:chargeErrorHistogram}).
The per-target underbinding shows the same picture: the three targets with exclusively charged ligands (\emph{mcl1}, \emph{bace}, and \emph{thrombin}) span \SI{3.4}{} to \SI{7.0}{\kilo\calorie\per\mole}, inside the \SI{2.7}{} to \SI{7.5}{\kilo\calorie\per\mole} covered by the targets with exclusively neutral ligands.
Within \emph{cmet} and \emph{eg5}, the two targets that contain charged and neutral ligands, the charged ligands underbind more strongly, by \SI{3.2}{} and \SI{0.9}{\kilo\calorie\per\mole}, respectively.
This difference is target-specific and consistent with the weak representation of charged interactions in Martini (\cref{sec:results:pocketProperties}), whereas an inaccuracy of the electrostatics treatment would shift all charged ligands systematically.

\subsection{Per-Target Binding Free Energies}
\label{si:results:detailedBindingFreeEnergies}

\begin{figure}[H]
    \centering
    \includegraphics[width=\columnwidth]{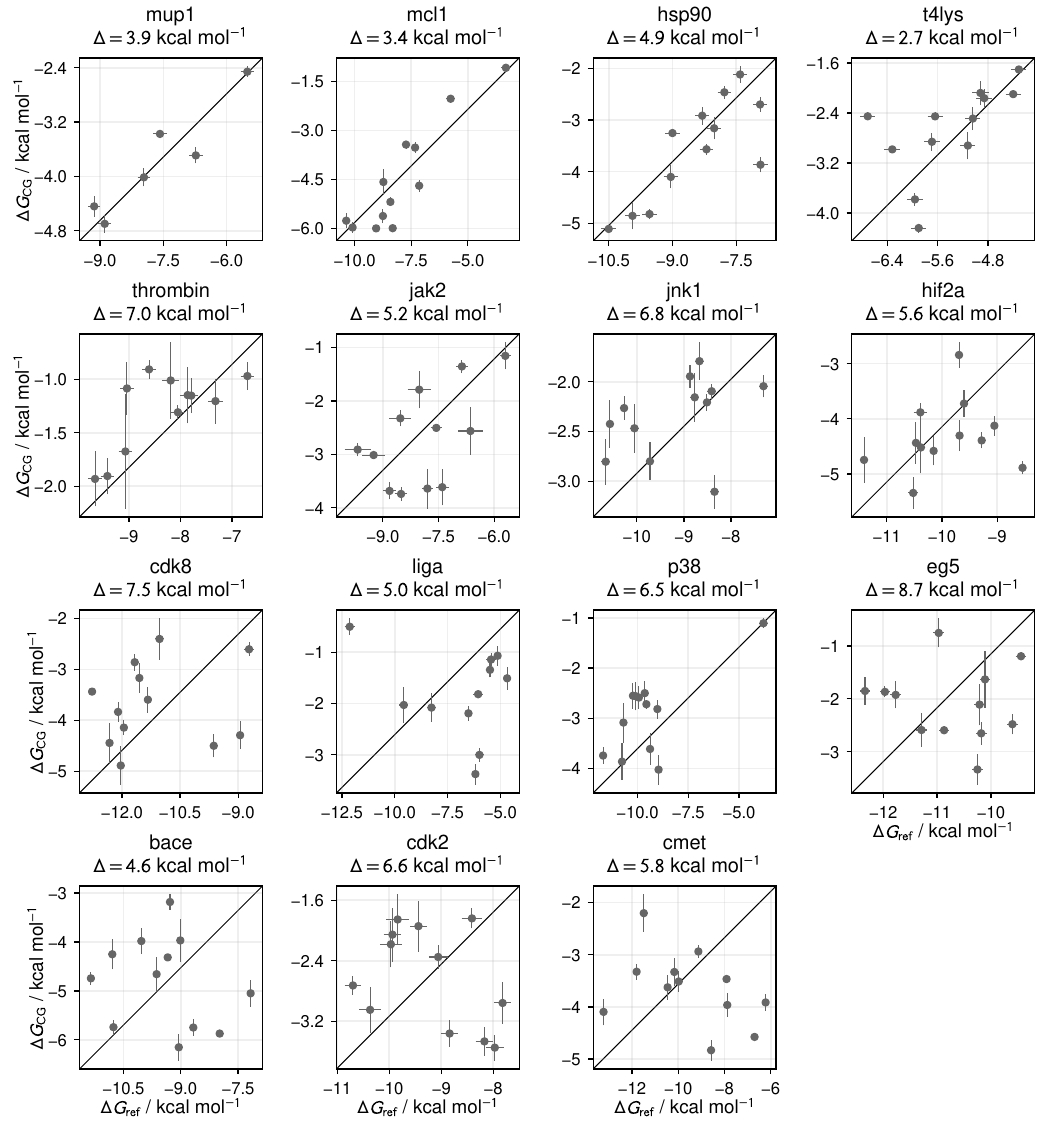}
    \caption{CG binding free energies $\Delta G_{\mathrm{CG}}$ against the FEP+ reference values $\Delta G_{\mathrm{ref}}$ of \citeauthor{Ross2023SchroedingerFEP}\cite{Ross2023SchroedingerFEP} for each of the 15 targets, ordered by decreasing ranking accuracy $\rho_{\mathrm{target}}$.
        The panel titles give the mean underbinding $\Delta = \langle \Delta G_{\mathrm{CG}} - \Delta G_{\mathrm{ref}} \rangle$ of the target.
        The diagonal marks the parity line shifted by $\Delta$ such that the spread around it reflects the ranking within the target.
        Error bars indicate one standard error.
        Note that axis ranges differ between the panels.}
    \label{fig:si:allTargetScatterplots}
\end{figure}

\begin{figure}[H]
    \centering
    \includegraphics[width=\columnwidth]{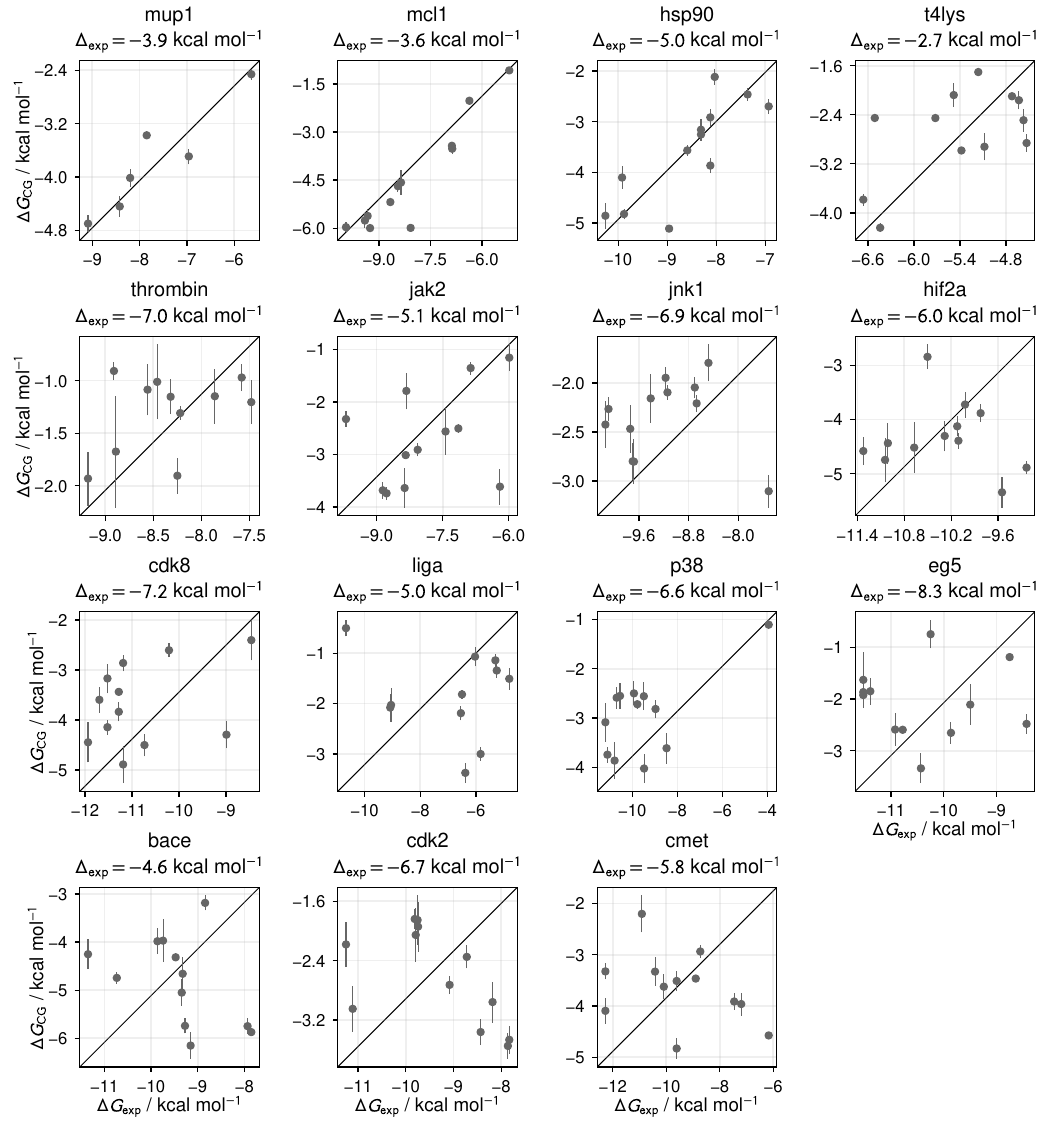}
    \caption{CG binding free energies $\Delta G_{\mathrm{CG}}$ against the experimental values $\Delta G_{\mathrm{exp}}$ for each of the 15 targets, ordered as in \cref{fig:si:allTargetScatterplots}, i.e., by decreasing ranking accuracy $\rho_{\mathrm{target}}$ against the FEP+ reference.
        The panel titles give the mean underbinding against the experimental reference $\Delta_{\mathrm{exp}} = \langle \Delta G_{\mathrm{CG}} - \Delta G_{\mathrm{exp}} \rangle$ of the target and the diagonal marks the parity line shifted by $\Delta_{\mathrm{exp}}$.
        Error bars indicate one standard error of the CG values; no uncertainties are reported for the experimental values.
        Note that axis ranges differ between the panels.}
    \label{fig:si:allTargetScatterplotsExp}
\end{figure}

\begin{figure}[H]
    \centering
    \includegraphics[width=\columnwidth]{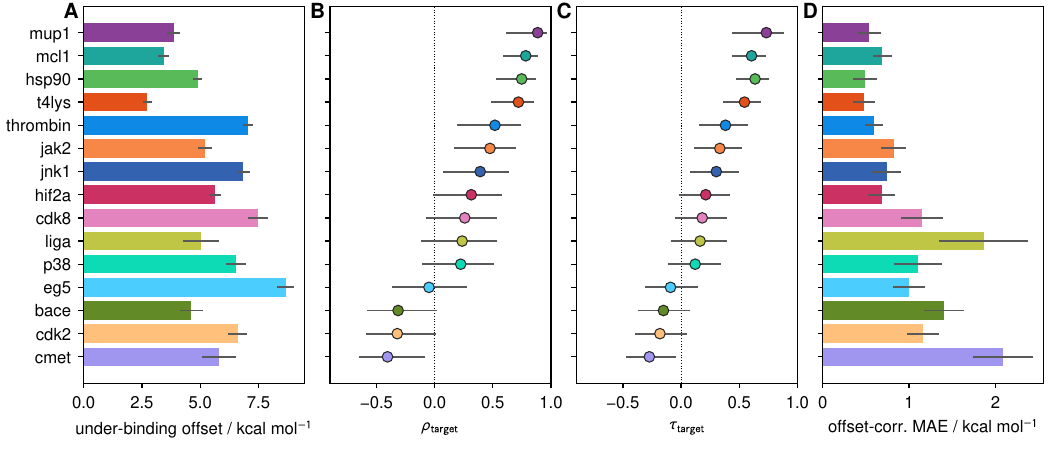}
    \caption{Per target summary statistics of the CG binding free energies against the FEP+ reference, ordered by decreasing ranking accuracy $\rho_{\mathrm{target}}$.
        \textbf{(A)} Mean underbinding $\Delta = \langle \Delta G_{\mathrm{CG}} - \Delta G_{\mathrm{ref}} \rangle$ of each target.
        \textbf{(B, C)} Spearman $\rho_{\mathrm{target}}$ and Kendall $\tau_{\mathrm{target}}$ rank correlation between the CG and the reference binding free energies.
        Both correlation metrics agree on the ranking accuracy per target.
        \textbf{(D)} MAE after removing the per target offset $\Delta$.
        Error bars give one standard error, computed as the standard error of the mean over the ligands of the target for $\Delta$ and the offset-corrected MAE, and with a Fisher $z$-transform for the two rank correlations.}
    \label{fig:si:freeEnergyStats}
\end{figure}

\subsection{Structure--Activity Relationships}
\label{si:results:structureActivityRelationships}

\begin{table}[H]
    \caption{Structure--activity relationships (SARs) of the ligand series of all 15 targets.
        $n$ is the number of ligands of the series and $\rho_{\mathrm{target}}$ the Spearman correlation between the CG and the reference binding free energies.
        Each remaining column gives the SAR of one ligand property, i.e., the Spearman correlation between the reference potency $-\Delta G_{\mathrm{ref}}$ and that property within the series (\cref{sec:methods:descriptors}).
        A positive value therefore indicates that ligands with a higher value of the respective property bind more strongly.
        Properties that do not vary within a series are reported as zero.
        The ligand logP is the mean of four predictions (\cref{sec:methods:ligands}) and the interaction counts are obtained with PLIP\cite{Salentin2015PLIP} on the atomistic input structures.}
    \label{tab:si:SARs}
    \input{si-ranking-data}
\end{table}

\begin{figure}[H]
    \centering
    \includegraphics[width=4in]{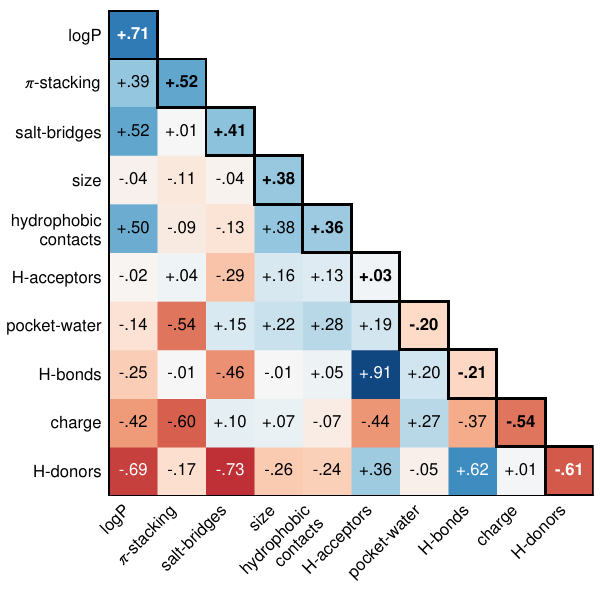}
    \caption{Full version of the SAR correlation matrix of \cref{fig:rankingModel}A, showing all structure--activity relationships (\cref{tab:si:SARs}) except the $\pi$-cation SAR, which is zero for every series.
        The diagonal entries (boxed and bold) give the Pearson correlation coefficient $r$ of each SAR with the ranking accuracy $\rho_{\mathrm{target}}$, the off-diagonal entries the cross-correlations between the SARs, both computed across the 15 targets.
        The color encodes $r$ from $-1$ (red) to $+1$ (blue).
        Since values are computed across only 15 targets and from a limited number of ligands per target, the coefficients are only indicative and not statistically significant.}
    \label{fig:si:rankingHeatmapFull}
\end{figure}

\subsection{Comparison with MMGBSA}
\label{si:results:mmgbsaComparison}

\begin{figure}[H]
    \centering
    \includegraphics[width=\columnwidth]{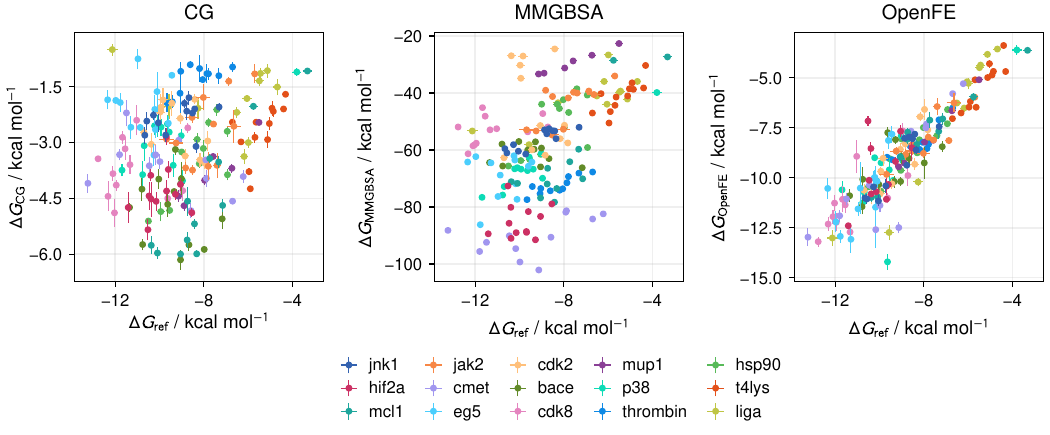}
    \caption{Absolute binding free energies of the CG, MMGBSA, and OpenFE calculations against the FEP+ reference values $\Delta G_{\mathrm{ref}}$ for all ligands.
        Error bars give one standard error; no uncertainties are reported for the MMGBSA values.
        Note the different axis ranges of the three panels: the CG values are systematically shifted towards weaker binding but cover a range comparable to the reference, whereas the MMGBSA values span a much wider range.
        The MMGBSA and OpenFE values are taken from \citeauthor{Baumann2026OpenFE} and the FEP+ reference from \citeauthor{Ross2023SchroedingerFEP}\cite{Ross2023SchroedingerFEP, Baumann2026OpenFE}}
    \label{fig:si:fullScatterComparison}
\end{figure}

\begin{figure}[H]
    \centering
    \includegraphics[width=\columnwidth]{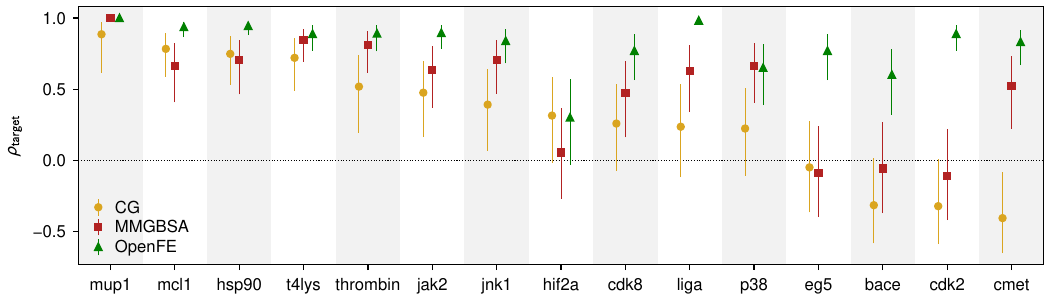}
    \caption{Ranking accuracy $\rho_{\mathrm{target}}$ of the CG, MMGBSA, and OpenFE results against the FEP+ reference, resolved by target, i.e., the target-wise version of \cref{fig:spearmanRhosMMGBSA}A.
        Targets are sorted by decreasing $\rho_{\mathrm{target}}$ of the CG results.
        Error bars give one standard error computed with a Fisher $z$-transform.
        Since each correlation is based on only 6 to 12 ligands, the individual values should only be interpreted qualitatively.}
    \label{fig:si:correlationComparison}
\end{figure}

\section{Ligand Topologies}
\label{si:ligandTopologies}
List of all ligands and their CG representations, sorted by target.
Structures are aligned by the shared substructure, making the congeniality of the series apparent.
Coloured overlays mark the atoms mapped onto each bead and are labelled with the bead type.
For charged beads, a neutral bead type is given as an alternative for a neutralized version of the ligand.
Dashed outlines mark virtual sites.

    {\footnotesize
        \tikz[baseline=-0.5ex]\filldraw[beadC,cgkey] (0,0) circle (1.7pt);~apolar (C)\quad
        \tikz[baseline=-0.5ex]\filldraw[beadN,cgkey] (0,0) circle (1.7pt);~intermediate (N)\quad
        \tikz[baseline=-0.5ex]\filldraw[beadP,cgkey] (0,0) circle (1.7pt);~polar (P)\quad
        \tikz[baseline=-0.5ex]\filldraw[beadQ,cgkey] (0,0) circle (1.7pt);~charged (Q)\quad
        \tikz[baseline=-0.5ex]\filldraw[beadX,cgkey] (0,0) circle (1.7pt);~halo-compound (X)\quad
        \tikz[baseline=-0.5ex]\filldraw[black!45,cgkey,cghulldash] (0,0) circle (1.7pt);~virtual site
    }

\begin{figure}[H]
    \centering
    \makebox[\textwidth][c]{\includegraphics{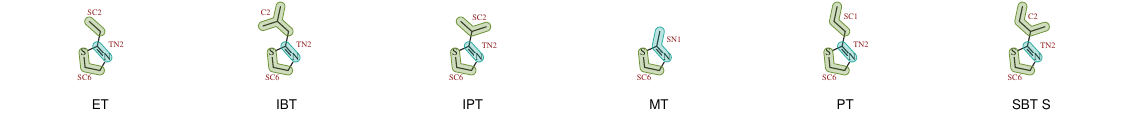}}
    \caption{Martini~3 CG mapping of the MUP-1 ligands.}
    \label{fig:cg:mup1}
\end{figure}

\begin{figure}[H]
    \centering
    \makebox[\textwidth][c]{\includegraphics{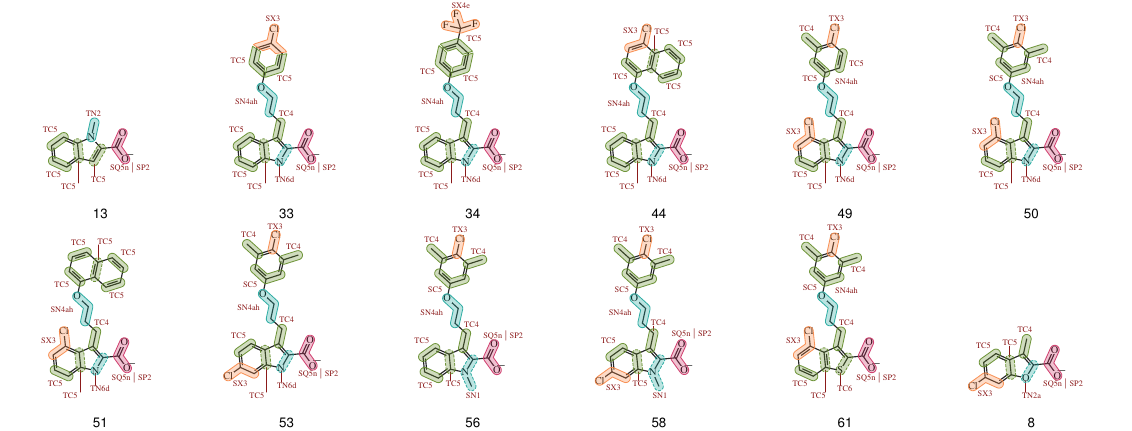}}
    \caption{Martini~3 CG mapping of the Mcl-1 ligands.}
    \label{fig:cg:mcl1}
\end{figure}

\begin{figure}[H]
    \centering
    \makebox[\textwidth][c]{\includegraphics{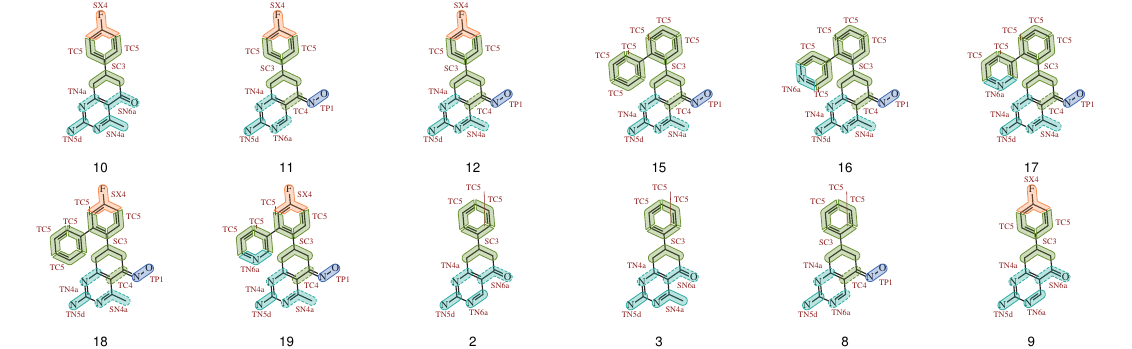}}
    \caption{Martini~3 CG mapping of the Hsp90 ligands.}
    \label{fig:cg:hsp90}
\end{figure}

\begin{figure}[H]
    \centering
    \makebox[\textwidth][c]{\includegraphics{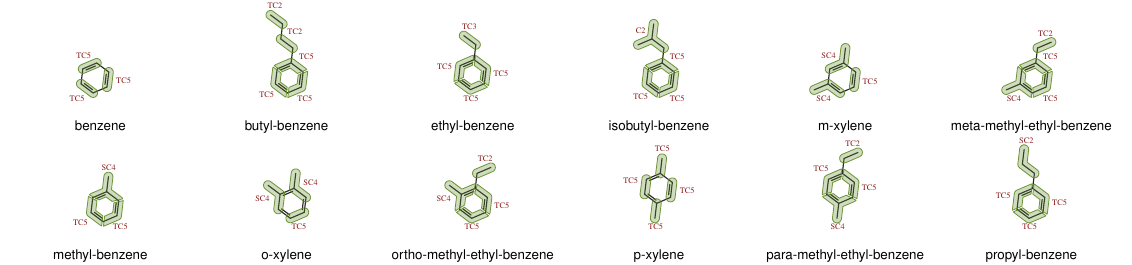}}
    \caption{Martini~3 CG mapping of the T4 lysozyme L99A ligands.}
    \label{fig:cg:t4lys}
\end{figure}

\begin{figure}[H]
    \centering
    \makebox[\textwidth][c]{\includegraphics{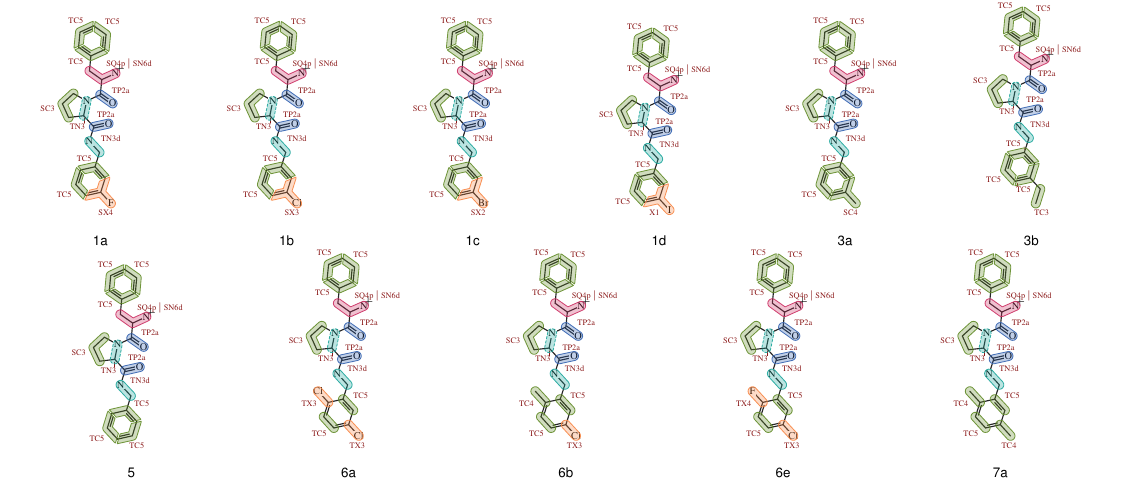}}
    \caption{Martini~3 CG mapping of the thrombin ligands.}
    \label{fig:cg:thrombin}
\end{figure}

\begin{figure}[H]
    \centering
    \makebox[\textwidth][c]{\includegraphics{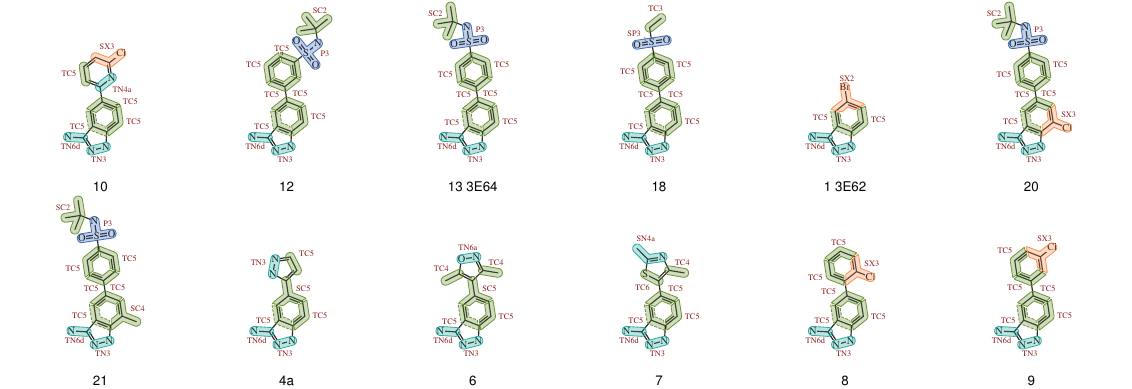}}
    \caption{Martini~3 CG mapping of the JAK2 ligands.}
    \label{fig:cg:jak2}
\end{figure}

\begin{figure}[H]
    \centering
    \makebox[\textwidth][c]{\includegraphics{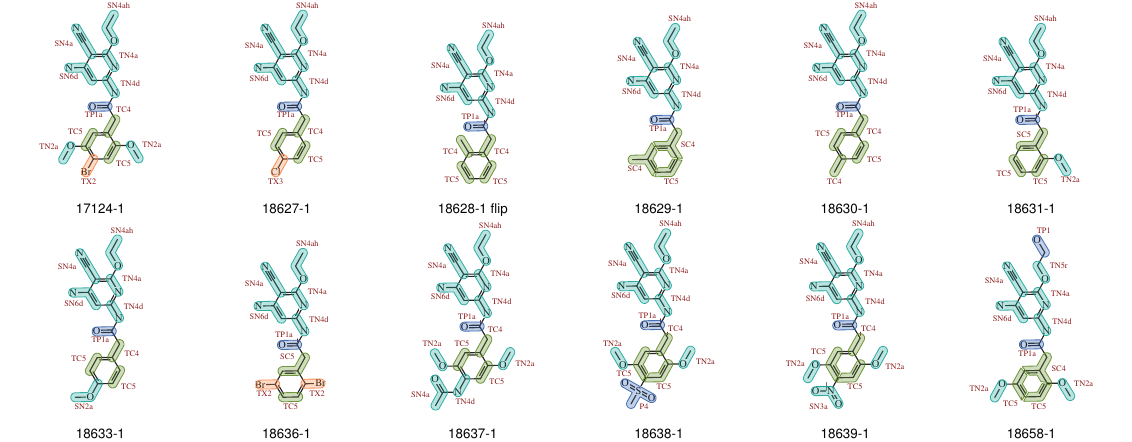}}
    \caption{Martini~3 CG mapping of the JNK1 ligands.}
    \label{fig:cg:jnk1}
\end{figure}

\begin{figure}[H]
    \centering
    \makebox[\textwidth][c]{\includegraphics{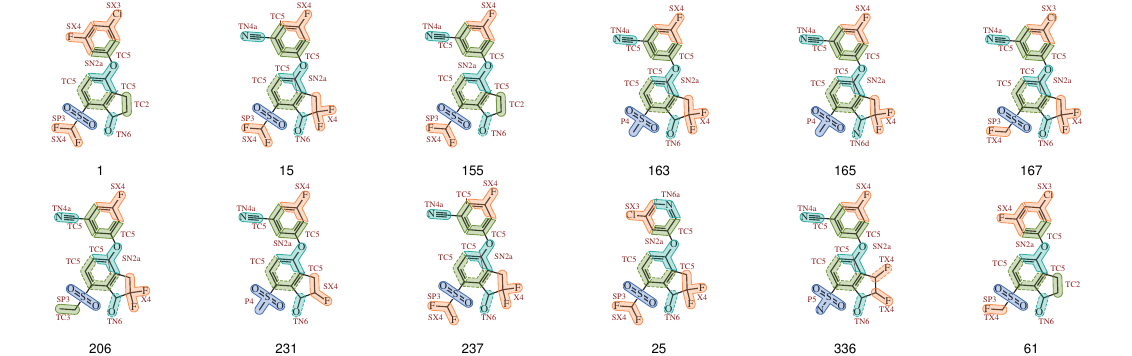}}
    \caption{Martini~3 CG mapping of the HIF-2$\alpha$ ligands.}
    \label{fig:cg:hif2a}
\end{figure}

\begin{figure}[H]
    \centering
    \makebox[\textwidth][c]{\includegraphics{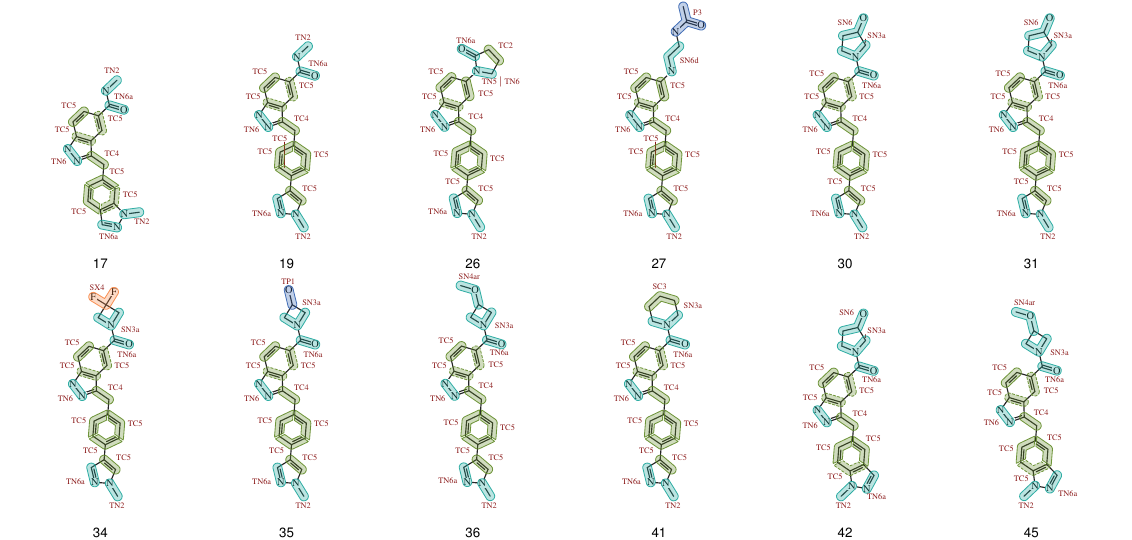}}
    \caption{Martini~3 CG mapping of the CDK8 ligands.}
    \label{fig:cg:cdk8}
\end{figure}

\begin{figure}[H]
    \centering
    \makebox[\textwidth][c]{\includegraphics{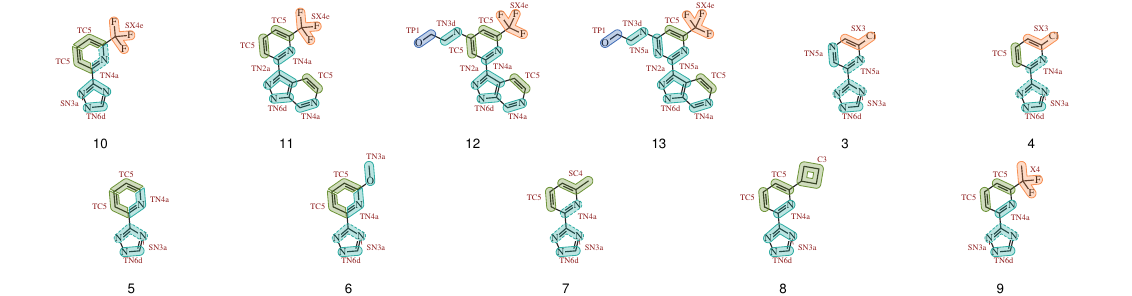}}
    \caption{Martini~3 CG mapping of the LigA ligands.}
    \label{fig:cg:liga}
\end{figure}

\begin{figure}[H]
    \centering
    \makebox[\textwidth][c]{\includegraphics{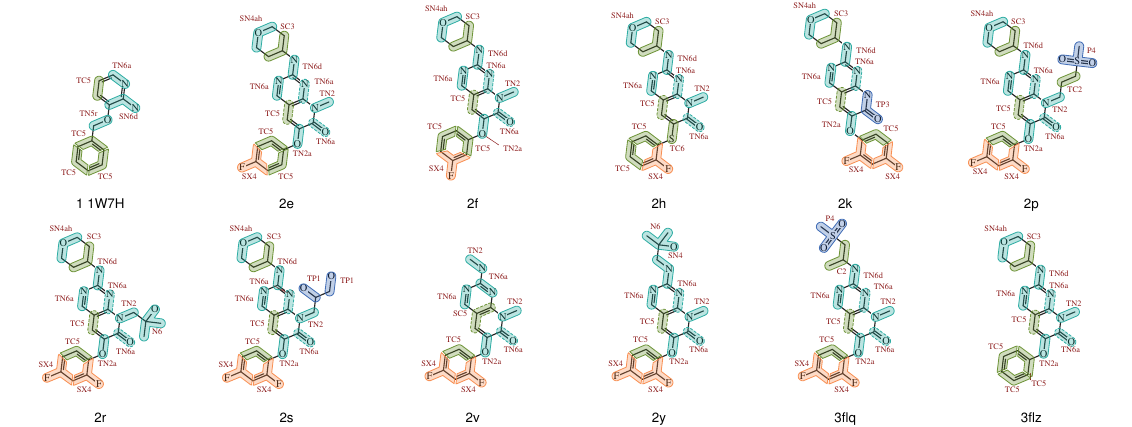}}
    \caption{Martini~3 CG mapping of the p38 ligands.}
    \label{fig:cg:p38}
\end{figure}

\begin{figure}[H]
    \centering
    \makebox[\textwidth][c]{\includegraphics{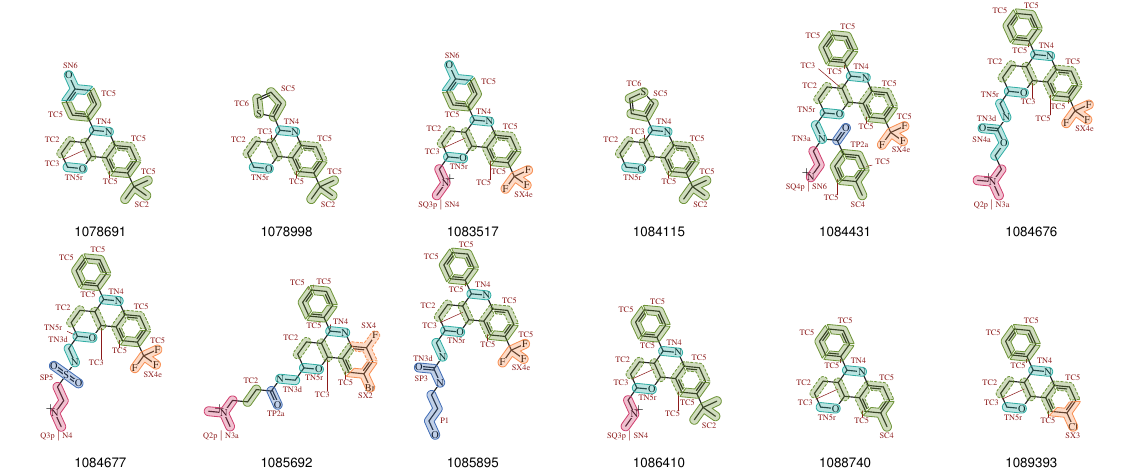}}
    \caption{Martini~3 CG mapping of the Eg5 ligands.}
    \label{fig:cg:eg5}
\end{figure}

\begin{figure}[H]
    \centering
    \makebox[\textwidth][c]{\includegraphics{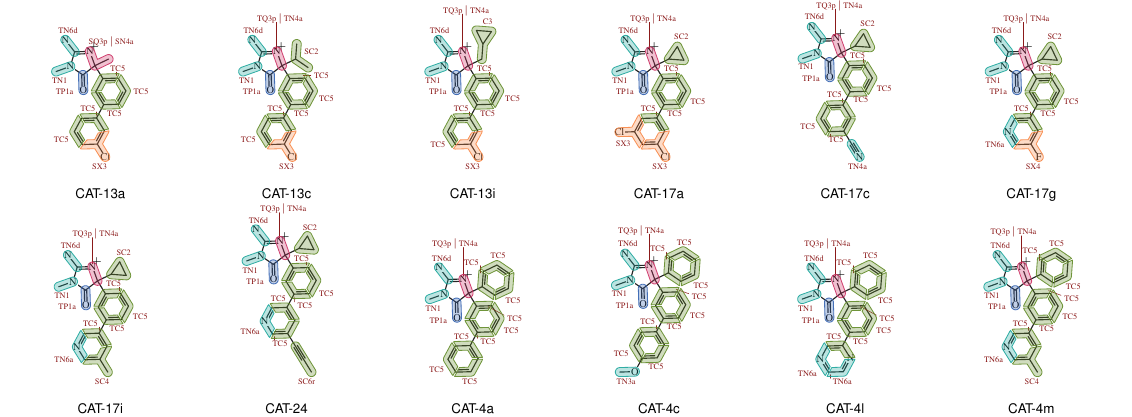}}
    \caption{Martini~3 CG mapping of the BACE1 ligands.}
    \label{fig:cg:bace}
\end{figure}

\begin{figure}[H]
    \centering
    \makebox[\textwidth][c]{\includegraphics{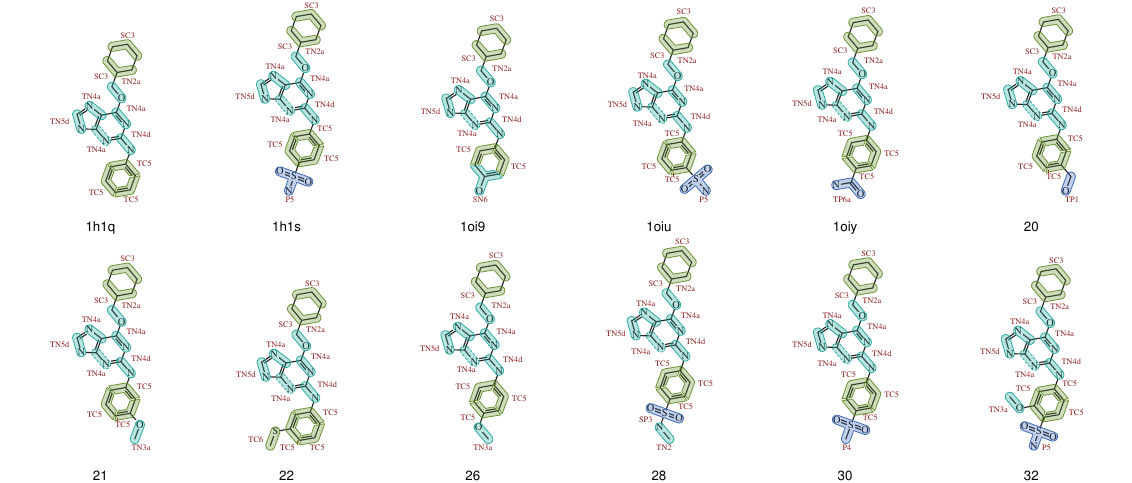}}
    \caption{Martini~3 CG mapping of the CDK2 ligands.}
    \label{fig:cg:cdk2}
\end{figure}

\begin{figure}[H]
    \centering
    \makebox[\textwidth][c]{\includegraphics{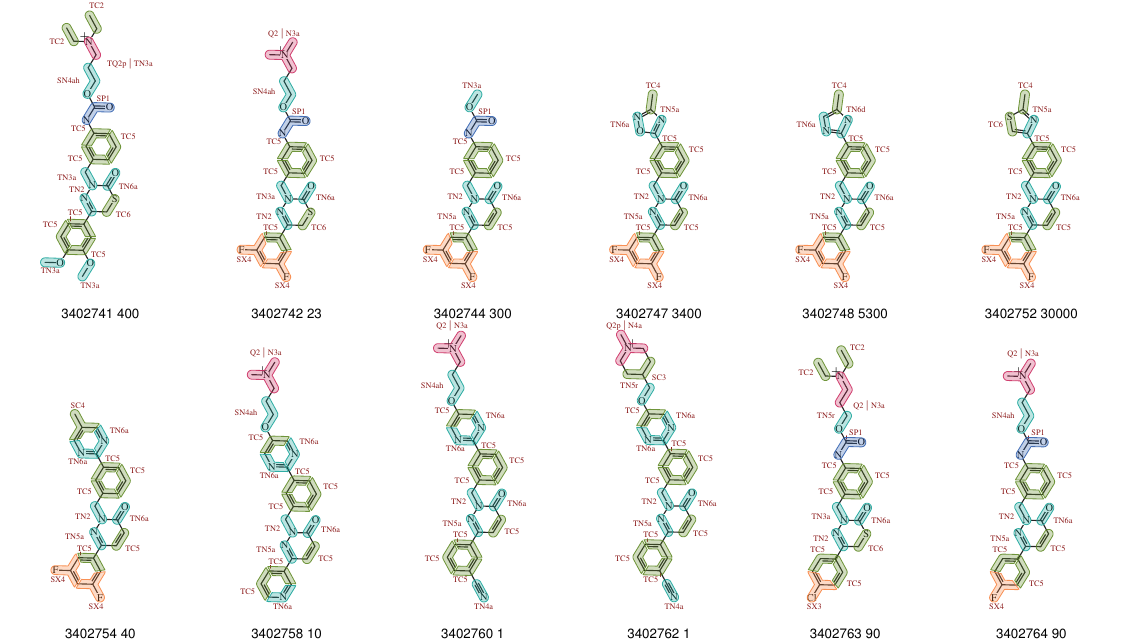}}
    \caption{Martini~3 CG mapping of the c-Met ligands.}
    \label{fig:cg:cmet}
\end{figure}

%% file: si-target-data.tex
\begin{tabular}{lcccccccccccc}
\toprule
 & & & \multicolumn{7}{c}{Protein pocket} & \multicolumn{3}{c}{Ligand series} \\
\cmidrule(lr){4-10} \cmidrule(lr){11-13}
 & \thead{$n$} & \thead{protein\\residues} & \thead{pocket\\residues} & \thead{charged\\residues} & \thead{$f_{\mathrm{charged}}$} & \thead{$f_{\mathrm{nonpolar}}$} & \thead{$f_{\mathrm{aromatic}}$} & \thead{$f_{\mathrm{polar}}$} & \thead{KD} & \thead{heavy\\atoms} & \thead{logP} & \thead{ligand\\charge} \\
\midrule
\textbf{mup1} & 6 & 156 & 16.5 & 0.0 & 0.00 & 0.79 & 0.32 & 0.21 & 2.02 & 6--9 & 1.0--2.3 & 0 \\
\textbf{mcl1} & 12 & 150 & 19.9 & 1.0 & 0.05 & 0.80 & 0.21 & 0.15 & 1.98 & 13--27 & 1.5--6.4 & -1 \\
\textbf{hsp90} & 12 & 208 & 21.7 & 3.0 & 0.14 & 0.64 & 0.11 & 0.22 & 0.57 & 18--27 & 1.7--3.8 & 0 \\
\textbf{t4lys} & 12 & 162 & 17.9 & 1.3 & 0.07 & 0.86 & 0.14 & 0.07 & 2.47 & 6--10 & 2.4--4.1 & 0 \\
\textbf{thrombin} & 11 & 296 & 26.8 & 7.0 & 0.26 & 0.45 & 0.25 & 0.29 & -0.58 & 26--28 & 2.0--3.2 & +1 \\
\textbf{jak2} & 12 & 292 & 23.2 & 5.4 & 0.23 & 0.57 & 0.09 & 0.20 & -0.26 & 11--25 & 1.4--3.4 & 0 \\
\textbf{jnk1} & 12 & 368 & 22.6 & 4.0 & 0.18 & 0.65 & 0.00 & 0.18 & 0.61 & 23--30 & 0.8--3.3 & 0 \\
\textbf{hif2a} & 12 & 110 & 36.7 & 2.5 & 0.07 & 0.47 & 0.24 & 0.46 & 0.49 & 24--28 & 1.7--3.6 & 0 \\
\textbf{cdk8} & 12 & 631 & 26.2 & 8.6 & 0.33 & 0.48 & 0.23 & 0.19 & -0.60 & 24--30 & 1.9--3.6 & 0 \\
\textbf{liga} & 11 & 308 & 17.1 & 4.1 & 0.24 & 0.50 & 0.11 & 0.26 & -0.19 & 11--23 & 0.6--2.4 & 0 \\
\textbf{p38} & 12 & 350 & 30.2 & 5.6 & 0.18 & 0.63 & 0.04 & 0.18 & 0.74 & 15--34 & 1.2--2.8 & 0 \\
\textbf{eg5} & 12 & 347 & 27.8 & 9.7 & 0.35 & 0.55 & 0.11 & 0.10 & -0.25 & 21--38 & 2.9--4.9 & 0/+1 \\
\textbf{bace} & 12 & 390 & 26.2 & 4.1 & 0.16 & 0.47 & 0.20 & 0.37 & -0.35 & 22--28 & 1.8--4.2 & +1 \\
\textbf{cdk2} & 12 & 555 & 22.6 & 7.2 & 0.32 & 0.51 & 0.13 & 0.18 & -0.43 & 24--30 & 2.2--4.3 & 0 \\
\textbf{cmet} & 12 & 285 & 26.7 & 6.4 & 0.24 & 0.50 & 0.14 & 0.26 & -0.44 & 27--37 & 2.0--4.4 & 0/+1 \\
\bottomrule
\end{tabular}

%% file: si-ranking-data.tex
\begin{tabular}{lccccccccccccccc}
\toprule
 & \thead{n} & \thead{$\rho_{\mathrm{target}}$} & \thead{logP} & \thead{size} & \thead{charge} & \thead{total H\\bonds} & \thead{H\\donors} & \thead{H\\acceptors} & \thead{salt-\\bridges} & \thead{pocket\\water} & \thead{hydrophobic\\contacts} & \thead{$\pi$\\stacking} & \thead{$\pi$-cation} \\
\midrule
\textbf{mup1} & 6 & 0.89 & 1.00 & 0.97 & 0.00 & 0.00 & 0.00 & 0.00 & 0.00 & 0.00 & 0.96 & 0.00 & 0.00 \\
\textbf{mcl1} & 12 & 0.78 & 0.63 & 0.49 & 0.00 & 0.00 & 0.00 & 0.00 & 0.00 & 0.00 & 0.61 & 0.25 & 0.00 \\
\textbf{hsp90} & 12 & 0.75 & 0.75 & 0.71 & 0.00 & 0.00 & 0.00 & 0.00 & 0.00 & 0.33 & 0.57 & 0.59 & 0.00 \\
\textbf{t4lys} & 12 & 0.72 & 0.71 & 0.81 & 0.00 & 0.00 & 0.00 & 0.00 & 0.00 & 0.00 & 0.80 & 0.00 & 0.00 \\
\textbf{thrombin} & 11 & 0.52 & 0.88 & 0.45 & 0.00 & 0.00 & 0.00 & 0.00 & 0.00 & 0.00 & 0.45 & 0.00 & 0.00 \\
\textbf{jak2} & 12 & 0.48 & 0.64 & 0.66 & 0.00 & 0.41 & 0.22 & 0.70 & 0.00 & 0.00 & 0.22 & 0.00 & 0.00 \\
\textbf{jnk1} & 12 & 0.39 & -0.49 & 0.84 & 0.00 & 0.13 & 0.13 & 0.13 & 0.00 & 0.00 & -0.62 & 0.00 & 0.00 \\
\textbf{hif2a} & 12 & 0.31 & 0.19 & 0.50 & 0.00 & 0.22 & 0.00 & 0.22 & 0.00 & 0.22 & -0.08 & 0.22 & 0.00 \\
\textbf{cdk8} & 12 & 0.26 & 0.11 & 0.77 & 0.00 & -0.18 & 0.13 & -0.25 & 0.00 & -0.13 & -0.13 & 0.25 & 0.00 \\
\textbf{liga} & 11 & 0.24 & 0.68 & 0.91 & 0.00 & 0.77 & 0.00 & 0.77 & 0.00 & 0.41 & 0.80 & 0.00 & 0.00 \\
\textbf{p38} & 12 & 0.22 & -0.17 & 0.74 & 0.00 & 0.65 & 0.39 & 0.48 & 0.00 & 0.88 & 0.48 & -0.48 & 0.00 \\
\textbf{eg5} & 12 & -0.05 & -0.50 & 0.53 & 0.39 & 0.40 & 0.48 & 0.14 & 0.00 & 0.00 & -0.20 & 0.00 & 0.00 \\
\textbf{bace} & 12 & -0.31 & 0.44 & -0.31 & 0.00 & 0.20 & 0.36 & -0.05 & 0.00 & 0.00 & -0.17 & 0.00 & 0.00 \\
\textbf{cdk2} & 12 & -0.32 & -0.87 & 0.67 & 0.00 & 0.77 & 0.82 & 0.50 & -0.22 & 0.00 & 0.54 & 0.00 & 0.00 \\
\textbf{cmet} & 12 & -0.41 & -0.33 & 0.78 & 0.66 & -0.45 & 0.00 & -0.45 & 0.00 & 0.55 & 0.48 & -0.71 & 0.00 \\
\bottomrule
\end{tabular}